\documentclass[a4paper, 12pt]{article}
\usepackage[english]{babel}

\usepackage[letterpaper,top=2cm,bottom=2cm,left=3cm,right=3cm,marginparwidth=1.75cm]{geometry}

\usepackage{amsmath}
\usepackage{amssymb}
\usepackage{amsthm}
\usepackage{thmtools}
\usepackage{thm-restate}
\usepackage{graphicx}
\usepackage[colorlinks=true, allcolors=blue]{hyperref}
\usepackage[capitalise]{cleveref}
\usepackage{xspace}
\usepackage[dvipsnames]{xcolor}
\usepackage{tikz}
\usepackage{enumerate}

\newtheorem{theorem}{Theorem}
\newtheorem{lemma}[theorem]{Lemma}
\newtheorem{observation}[theorem]{Observation}
\newtheorem{definition}[theorem]{Definition}
\newtheorem{claim}[theorem]{Claim}

\newenvironment{claimproof}[1][\proofname]{
  
  \begin{proof}[#1]}{\end{proof}}

\newcommand{\pls}{\text{PLS}\xspace}
\newcommand{\kopt}{\text{k-Opt}\xspace}
\newcommand{\infedge}{\text{non-edge}\xspace}
\newcommand{\fedge}{\text{$G$-edge}\xspace}
\newcommand{\fedges}{\text{$G$-edges}\xspace}

\newcommand{\R}{\mathbb{R}}

\title{The \kopt algorithm for the Traveling Salesman Problem has exponential running time for $k \geq 5$}
\author{Sophia Heimann\thanks{Research Institute for Discrete Mathematics, University of Bonn, Germany (s6soheim@uni-bonn.de)}, 
        Hung P. Hoang\thanks{Algorithms and Complexity Group, Faculty of Informatics, TU Wien, Austria, (phoang@ac.tuwien.ac.at) funded by the Austrian Science Foundation (FWF, project Y1329 START-Programm)}, 
        Stefan Hougardy\thanks{Research Institute for Discrete Mathematics and Hausdorff Center for Mathematics, University of Bonn, Germany (hougardy@dm.uni-bonn.de) funded by the Deutsche Forschungsgemeinschaft (DFG, German Research Foundation) under Germany's Excellence Strategy -- EXC-2047/1 -- 390685813}}

\begin{document}
\maketitle

\begin{abstract}
 The \kopt algorithm is a local search algorithm for the Traveling Salesman Problem.
 Starting with an initial tour, it iteratively replaces at most $k$ edges in the tour with the same number of edges to obtain a better tour.
 Krentel (FOCS 1989) showed that the Traveling Salesman Problem with the \kopt neighborhood is complete for the class PLS (polynomial time local search) and that the \kopt algorithm can have exponential running time for any pivot rule.
 However, his proof requires $k \gg 1000$ and has a substantial gap.
 We show the two properties above for a much smaller value of $k$, addressing an open question by Monien, Dumrauf, and Tscheuschner (ICALP 2010).
 In particular, we prove the \pls-completeness for $k \geq 17$ and the exponential running time for $k \geq 5$.
 \end{abstract}

\section{Introduction}

The well-known Traveling Salesman Problem (TSP) consists of finding a spanning cycle of an edge weighted complete graph, such that the total edge weight of the cycle is the smallest possible.
A popular heuristic for this problem is a local search algorithm called \kopt.
Starting with an arbitrary tour, it iteratively replaces at most $k$ edges in the tour with the same number of edges, as long as the resulting tour has smaller total edge weight.
We define TSP/\kopt to be the problem of finding a local optimum for a TSP instance 
with the \kopt algorithm.

A fundamental question in the area of local search algorithms is to determine the number of iterations a given
local search algorithm may need in the worst case. A local search algorithm with a specified pivot rule 
has the \emph{is-exp} property if there exist problem instances and initial solutions for which 
the local search algorithm requires an exponential number of iterations. For example, it is well known that 
the Simplex algorithm for linear programming has the is-exp property
for many different pivot rules~\cite{KleeMinty,JEROSLOW1973367,Avis1978,GOLDFARB1979277}. 
For TSP, Chandra, Karloff, and Tovey~\cite{Chandra1999} showed that 
TSP/\kopt has the is-exp property. This even holds for Euclidean TSP with the 2-Opt neighborhood~\cite{Englert2014}.

For the Simplex algorithm it is not known whether there exists a pivot rule that guarantees a polynomial number 
of iterations. This is one of the most important open problems in the area of linear programming. 
Contrary to this Krentel~\cite{Kre1989} proved in 1989 that for sufficiently large values of $k$, TSP/\kopt exhibits the \emph{all-exp} property, that is, the \kopt algorithm requires an exponential number of iterations to find a local optimum, for all possible pivot rules and for infinitely many pairs of a TSP instance and an initial tour. 
Krentel estimated that his proof yields a value for $k$ between $1{,}000$ and $10{,}000$. 
By using a straight forward way to implement some missing details in Krentel's proof it was recently shown that 
his proof yields the value $14{,}208$ for $k$~\cite{HH2023}. 

Following Krentel's paper there have been claims in other papers~\cite{JG1997, Yan1997} through private communication with Krentel that a careful analysis of the original proof can bring down the value to $k = 8$ 
and conceivably to $k = 6$. However, there has been no available written proof for these claims.
In fact, up to date, the 1989 paper of Krentel~\cite{Kre1989} is the only paper on the topic.
Consequently, Monien, Dumrauf, and Tscheuschner~\cite{Monien_Dumrauf_Tscheuschner_2010} posed an open question on the complexity of TSP/\kopt for $k \ll 1000$.

In this paper, we show that TSP/\kopt has the all-exp property already for much smaller values of $k$:
\begin{theorem}
\label{thm:all_exp_5}
    TSP/\kopt has the all-exp property for $k \geq 5$.
\end{theorem}

Our proof of \cref{thm:all_exp_5} is based on a new reduction from the bounded degree Max-Cut problem to TSP (see~\cref{sec:reduction}) which involves the construction of so called \emph{parity gadgets} (see~\cref{sec:parity-gadget}). 
With a first such approach we are able to prove the all-exp property of TSP/\kopt for $k\ge 13$ (see \cref{sec:all-exp-13}).
To lower the value of $k$ additional ideas are required. First, we exploit the structure of a recent construction
of Michel and Scott~\cite{Michel_Max_Cut_4} of a degree-four bounded Max-Cut instance with the all-exp property under the flip neighborhood. 
Second, we show how to
use global properties of our overall reduction to relax some local conditions on our parity gadgets. 
Combining these two ideas we achieve the value $k \geq 9$ (see \cref{sec:proof_all_exp_9}).    
To arrive at our final result for $k \geq 5$ we have to modify the construction of Michel and Scott~\cite{Michel_Max_Cut_4}. 
Moreover, we have to combine our parity gadgets with so called \emph{double gadgets} and use a labeling scheme to 
assign different gadgets at different places in the reduction. These results we present in~\cref{sec:all-exp-5}.

The second main contribution of our paper is a proof of the following result:

\begin{restatable}{theorem}{plscomplete}
\label{thm:PLS_complete_17}
    TSP/\kopt is PLS-complete for $k \geq 17$.
\end{restatable}

The complexity class PLS and the notion of PLS-completeness (for definitions see \cref{sec:preliminaries})
were introduced in 1988 by Johnson, Papdimitriou, and Yannakakis~\cite{JPY1988} 
to capture the observation that for many NP-hard problems it is not only difficult to compute a global optimum but
even computing a local optimum is also hard. Examples of such problems are the Maximum Satisfiability problem~\cite{krentel1990}, 
Max-Cut~\cite{schaffer1991}, and Set Cover~\cite{DS10}. The PLS-completeness of a problem means that a polynomial time
algorithm to find a local optimum for that problem would imply polynomial time algorithms for finding a local optimum for all problems in PLS. 

The PLS-completeness of TSP/\kopt was proved by Krentel~\cite{Kre1989} for $k \gg 1000$. However, his proof has a substantial gap 
as he assumes that edges of weight infinity cannot occur in a local optimum. We present in \cref{sec:PLS-complete-proof}
the first rigorous proof for the PLS-completeness of TSP/\kopt and at the same time drastically lower the value of
$k$ from Krentel's  $k \gg 1000$~\cite{Kre1989} to $k \ge 17$. Our proof uses several of the ideas used in our proof for 
\cref{{thm:all_exp_5}}. But in this case we need to take more care on the order in which the parity gadgets are plugged together in our construction. In addition, we show in~\cref{lem:no_infty_local_optima} how to assign specific weights to the non-edges in our construction to prove that
no local optimum can contain such an edge. We achieve this by defining a weight assignment that exploits the special 
structure of the TSP instance resulting from our PLS-reduction.  
This is the first rigorous proof of such a result for the \kopt algorithm and there seems 
not to be a generic way to prove it for arbitrary TSP instances (as for example those constructed by Krentel~\cite{Kre1989}).

\section{Preliminaries}
\label{sec:preliminaries}

\subsection{Local search problems and the class \pls}
A \emph{local search problem} $P$ is an optimization problem that consists of a set of instances $D_{P}$, a finite set of (feasible) solutions $F_{P}(I)$ for each instance $I\in D_{P}$, an objective function $f_{P}$ that assigns an integer value to each instance $I\in D_{P}$ and solution $s\in F_P(I)$, and a neighborhood $N_{P}(s,I)\subseteq F_{P}(I)$ for each solution $s\in F_{P}(I)$. 
The size of every solution $s \in F_{P}(I)$ is bounded by a polynomial in the size of $I$. 
The goal is to find a \emph{locally optimal solution} for a given instance $I$; that is, a solution $s \in F_{P}(I)$, such that no solution $s' \in N_{P}(s,I)$ yields a better objective value than $f_P(s,I)$.
Formally, this means, for all $s'\in N_{P}(s,I)$, $f_{P}(s,I)\leq f_{P}(s',I)$ if $P$ is a minimization problem, and $f_{P}(s,I)\geq f_{P}(s',I)$ if $P$ is a maximization problem.

A \emph{standard local search algorithm} for an instance $I$ proceeds as follows.
It starts with some initial solution $s \in F_{P}(I)$.
Then it iteratively visits a neighbor with better objective value, until it reaches a local optimum. 
If a solution has more than one better neighbor, the algorithm has to choose one by some prespecified rule, often referred as a \emph{pivot rule}.

\begin{definition}[All-exp property]
    A local search problem~$P$ has the \emph{all-exp} property, if there are infinitely many pairs of an instance~$I$ of $D_P$ and an initial solution $s \in F_P(I)$, for which the standard local search algorithm always needs an exponential number of iterations 
    for all possible pivot rules. 
\end{definition}

\begin{definition}[The class~\pls~\cite{JPY1988}]
	A local search problem $P$ is in the class \emph{\pls}, if there are three polynomial time algorithms $A_{P}, \ B_{P}, \ C_{P}$ such that 
	\begin{itemize}
		\item Given an instance $I \in D_P$ , $A_{P}$ returns a solution  $s \in F_{P}(I)$;
		\item Given an instance $I \in D_{P}$ and a solution $s\in  F_{P}(I)$, $B_{P}$ computes the objective value $f_{P}(s,I)$ of $s$; and 
		\item Given an instance $I \in D_P$ and a solution $s \in F_{P}(I)$, $C_{P}$ returns a neighbor of $s$ with strictly better objective value, if it exists, and ``locally optimal", otherwise.
	\end{itemize}
\end{definition}

\begin{definition}[\pls-completeness~\cite{JPY1988}]
\label{def:pls_complete}
	A \emph{PLS-reduction} from a problem $P \in \pls$ to a problem $Q \in \pls$ is a pair of polynomial-time computable functions $h$ and $g$ that satisfy:
	\begin{enumerate}
		\item Given an instance $I \in D_{P}$, $h$ computes an instance $h(I) \in D_{Q}$; and
		\item Given an instance $I \in D_{P}$ and a solution $s_q \in F_{Q}(h(I))$, $g$ returns a solution $s_p \in F_{P}(I)$ such that if $s_q$ is a local optimum for $h(I)$, then $s_p$ is a local optimum for $I$. 
	\end{enumerate} 
	A problem $Q\in \pls$ is \emph{\pls-complete} if for every problem~$P \in \pls$, there exists a \pls reduction from $P$ to $Q$.
\end{definition}

\subsection{TSP/\kopt}
A \emph{spanning cycle}, a \emph{Hamiltonian cycle}, or a \emph{tour} of an undirected graph is a cycle that contains all vertices of the graph.

A TSP instance is a tuple $(G, w)$, where $G$ is a complete undirected graph $(V, E)$, and $w:E\to \R_{\geq 0}$ is a function that assigns a nonnegative weight to each edge of $G$.
The goal is to find a tour of $G$ that minimizes the sum of edge weights in the tour.
The definition of the class PLS requires that we have a polynomial time algorithm to find \emph{some}
solution. For complete graphs such an algorithm certainly exists. If the graph is not complete then 
because of the NP-completeness of the Hamiltonian cycle problem we do not know such an algorithm. 

A \emph{swap} is a tuple $(E_1, E_2)$ of subsets $E_1, E_2 \subseteq E$, $|E_1| = |E_2|$.
We say that it is a swap of $|E_1|$ edges.
If $|E_1| \leq k$ for some $k$, then we call it a $k$-swap.
Performing a swap $(E_1, E_2)$ from a subgraph $G'$ of $G$ refers to the act of removing $E_1$ from $G'$ and adding $E_2$ to $G'$.
We also call it swapping $E_1$ for $E_2$ in $G'$.
Given a tour~$\tau$, a swap $(E_1, E_2)$ is \emph{improving} for $\tau$, if after swapping $E_1$ for $E_2$ in $\tau$, we obtain a tour with lower total edge weight.

A \emph{($k$-)swap sequence} is a sequence $L = (S_1, \dots, S_{\ell})$, such that each $S_i$ is a ($k$-)swap.
For a tour $\tau$, we denote by $\tau^L$ the subgraph obtained from $\tau$ by performing $S_1, \dots, S_{\ell}$ in their order in $L$.
$L$ is \emph{improving} for a tour~$\tau$ if each $S_i$ is an improving ($k$-)swap for $\tau^{(S_1, \ldots, S_{i-1})}$.

The local search problem TSP/\kopt corresponds to TSP with the \kopt neighborhood (that is, the neighbors of a tour~$\tau$ are those that can be obtained from~$\tau$ by an improving $k$-swap).
The \kopt algorithm is then the standard local search algorithm for this problem, and an execution of the algorithm corresponds to an improving $k$-swap sequence.

\subsection{Max-Cut/Flip}
A Max-Cut instance is a tuple $(G, w)$, where $G$ is an undirected graph $(V,E)$ and $w: E \to \R$ is a function assigning weights to the edges of $G$.
A \emph{cut} $(V_1, V_2)$ of $G$ is a partition of the vertices of $G$ into two disjoint sets $V_1$ and $V_2$.
The \emph{cut-set} of a cut $(V_1, V_2)$ is the set of edges~$xy \in E$ such that $x \in V_1$ and $y \in V_2$.
The goal of Max-Cut is to find the cut to maximize the \emph{value} of the cut, that is the total weight of the edges in the cut-set.

Given a Max-Cut instance and an initial cut, the \emph{flip} of a vertex is a move of that vertex from a set of the cut to the other.
The flip of a vertex is \emph{improving}, if it results in an increase in the value of the cut.
For a cut~$\sigma$, its \emph{flip neighborhood} is the set of all cuts obtained from~$\sigma$ by an improving flip.
The Max-Cut/Flip problem is the local search problem that corresponds to the Max-Cut problem with the flip neighborhood.
We call its standard local search algorithm the \emph{Flip algorithm}.
A \emph{flip sequence} is a sequence $(v_1, \dots, v_{\ell})$ of vertices of $G$.
A flip sequence is \emph{improving}, if flipping the vertices in the order in the sequence increases the value of the cut at every step.
In other words, an improving flip sequence corresponds to an execution of the Flip algorithm.

Monien and Tscheuschner~\cite{Monien_Max_Cut_4} showed the all-exp property for Max-Cut/Flip even for graphs with bounded degree. 

\begin{theorem}[\cite{Monien_Max_Cut_4, Michel_Max_Cut_4}]
\label{thm:all_exp_maxcut}
    Max-Cut/Flip has the all-exp property, even when restricted to instances where all vertices have degree at most four.
\end{theorem}

Michel and Scott~\cite{Michel_Max_Cut_4} recently presented an alternative proof for \cref{thm:all_exp_maxcut}.
Interestingly, their construction is highly structured and exhibits a unique property: With a suitable initial cut, there is exactly one maximal improving flip sequence, and this sequence has exponential length.
We rely on this particular construction and especially the unique property to achieve the low value of $k$ in \cref{thm:all_exp_5}.

Note that \cref{thm:all_exp_maxcut} is tight with respect to the maximum degree, since the Flip algorithm on graphs with maximum degree at most three always terminates after a polynomial number of iterations~\cite{Poljak1995}.

\section{The main reduction}
\label{sec:reduction}
In this section, we describe the main reduction to TSP/\kopt from Max-Cut/Flip.

Let $(H, w)$ be a Max-Cut instance.
In order to avoid confusion with the vertices and edges in the TSP instance later on, we use \emph{$H$-vertices} and \emph{$H$-edges} for the vertices and edges of $H$.
We denote by $n$ and $m$ the number of $H$-vertices and $H$-edges, respectively.

We construct from $H$ the corresponding TSP instance as follows.
We start with a cycle of $3(n+m)$ edges.
We assign $n+m$ edges of this cycle to each of the $n$ $H$-vertices and the $m$ $H$-edges, such that any two assigned edges have distance at least two on the cycle.

\begin{figure}[ht]
    \centering
    \includegraphics{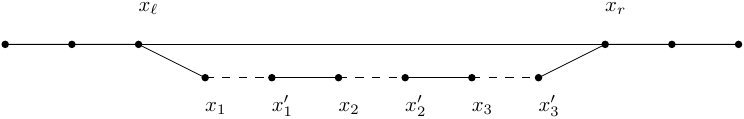}
    \caption{The first-set edge $x_{\ell} x_r$ and the second-set path $(x_{\ell}, x_1, x'_1, x_2, x'_2, x_3, x'_3, x_r)$ of an $H$-vertex $x$ of degree three. The dashed edges are gateways. The other edges of the second-set path are doors.}
    \label{fig:vertex_gadget}
\end{figure}

Next, in the cycle consider an edge that is assigned to an $H$-vertex $x$.
(Refer to \cref{fig:vertex_gadget} for an illustration of the following concepts.)
We label the two incident vertices of this edge $x_{\ell}$ and $x_r$, representing the left and the right vertex of the edge.
Let $d(x)$ be the degree of $x$ in $H$.
We add a new path of length $2d(x)+1$ to connect $x_{\ell}$ and $x_r$.
We call this new path the \emph{second-set path} of $x$, while we call the original edge that was assigned to $x$ the \emph{first-set edge} of $x$.
The idea is that the tour can connect $x_{\ell}$ and $x_r$ either via the first-set edge or via the second-set path.
This simulates whether the $H$-vertex $x$ is in the first set or second set of the cut for the Max-Cut problem.
Let $x_{\ell}, x_{1}, x'_{1}, \dots, x_{d(x)}, x'_{d(x)}, x_r$ be the labels of the vertices along the second-set path.
For $i \in \{1, \dots, d(x)\}$, we call the edge $x_i x'_i$ a \emph{gateway} of $x$.
The other edges of the second-set path are called the \emph{doors} of $x$.
In other words, we have alternating doors and gateways along the path, with doors at both ends of the path.

For each $H$-edge $xy$, we call the edge in the cycle of length $3(n+m)$  assigned to $xy$ the 
\emph{$xy$-edge}.
We remove a gateway of $x$, a gateway of $y$, and the $xy$-edge, and we connect the six incident vertices of the three removed edges by a \emph{parity gadget}. 

The purpose of this parity gadget is to simulate the contribution of the weight of edge $xy$ to the objective of the Max-Cut problem, based on whether $x$ and $y$ are in the same set. We will formally define the parity gadget 
in \cref{sec:parity-gadget}.

Finally, for each $H$-vertex~$x$, we assign an \emph{XOR gadget} to the first-set edge of~$x$ and the door of $x$ incident to $x_r$.
The purpose of the XOR gadget is to make sure that we can simulate only one flip in $H$ by a $k$-swap in the new graph.
The formal definition of the XOR gadget and its assignment are discussed in \cref{sec:xor_gadgets}.

\begin{figure}[ht]
    \centering
    \includegraphics[width=\hsize]{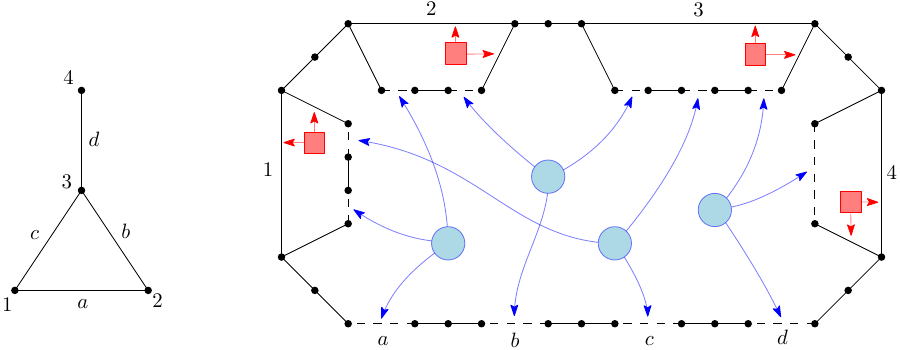}
    \caption{An example of our reduction from a Max-Cut instance (left figure) to a TSP instance (right figure). The parity gadgets are indicated by the blue circles attached to three edges each.
    The XOR gadgets are indicated by red boxes attached to two edges each.}
    \label{fig:reduction}
\end{figure}

Let $G$ be the resulting graph after all the operations above (see \cref{fig:reduction} for an example).
Except for certain edges in the parity gadgets, which we will specify later, the other edges have weight zero, including the edges in the XOR gadgets, the initial cycle, and the doors.
As a TSP instance requires a complete graph, we add the remaining edges with weight $\infty$ to obtain the final graph~$G_{\infty}$.
However, if we start with a tour with a finite total weight, the \kopt algorithm will never visit a tour that uses an edge with weight~$\infty$.
Hence, for the remaining of the reduction, we will argue based only on~$G$.

\subsection{Parity gadgets}
\label{sec:parity-gadget}
In this section, we specify the parity gadgets, formally defined as follows.

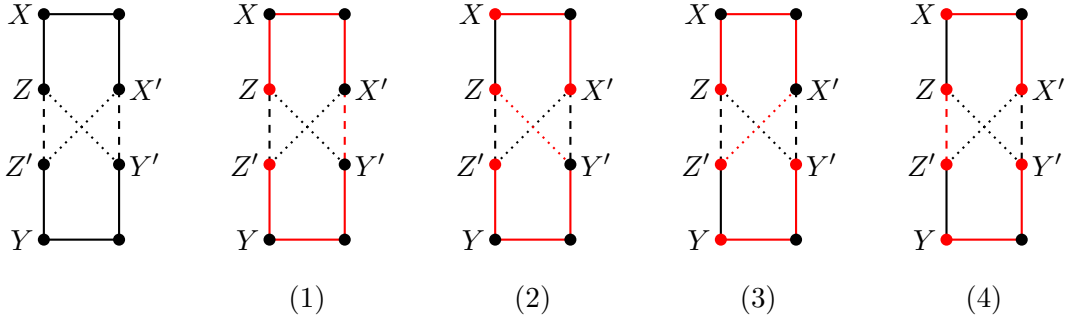
\begin{figure}[ht]
    \centering

\begin{tikzpicture}[scale=1.0]
\small
\def\defvertices{
\coordinate[label=left: $X$] (X) at (0,3);
\coordinate[label=left: $Y$] (Y) at (0,0);
\coordinate[label=left: $Z$] (Z) at (0,2);
\coordinate[label=right: $X'$] (X') at (1,2);
\coordinate[label=right: $Y'$] (Y') at (1,1);
\coordinate[label=left: $Z'$] (Z') at (0,1);
\coordinate[] (R) at (1,3);
\coordinate[] (S) at (1,0);
}
\def\drawvertices{
\fill[] (R) circle (0.8mm);
\fill[] (S) circle (0.8mm);}

\defvertices
\draw[thick] (Y') -- (S) -- (Y) -- (Z') (Z) -- (X) -- (R) -- (X');
\draw[thick, dashed] (Z') -- (Z)  (X') -- (Y');
\draw[thick, dotted] (Y') -- (Z) (X')--(Z');
\drawvertices
\fill[] (X) circle (0.8mm);
\fill[] (X') circle (0.8mm);
\fill[] (Y) circle (0.8mm);
\fill[] (Y') circle (0.8mm);
\fill[] (Z) circle (0.8mm);
\fill[] (Z') circle (0.8mm);

\begin{scope}[shift={(3,0)}]
\defvertices
\draw[thick] (Y')  (S)  (Y) (Z') (Z) (X) (R) (X');
\draw[thick, dotted] (Y') -- (Z) (X')--(Z');
\draw[thick, dashed] (Z') -- (Z);
\draw[thick, red] (Z) -- (X)--(R)--(X') (Y')--(S)--(Y)--(Z');
\draw[thick, dashed, red] (X') -- (Y');
\drawvertices
\draw (0.5, -0.8) node {(1)};
\fill[] (X) circle (0.8mm);
\fill[] (X') circle (0.8mm);
\fill[] (Y) circle (0.8mm);
\fill[] (Y') circle (0.8mm);
\fill[red] (Z) circle (0.8mm);
\fill[red] (Z') circle (0.8mm);
\end{scope}

\begin{scope}[shift={(6,0)}]
\defvertices
\draw[thick] (X) -- (Z) (Z')  (Y') (X');
\draw[thick, dotted] (X')--(Z');
\draw[thick, dashed] (Z') -- (Z)  (X') -- (Y');
\draw[thick, red] (Y')--(S)--(Y)--(Z') (X)--(R)--(X');
\draw[thick, dotted, red] (Y') -- (Z);
\drawvertices
\draw (0.5, -0.8) node {(2)};
\fill[red] (X) circle (0.8mm);
\fill[red] (X') circle (0.8mm);
\fill[] (Y) circle (0.8mm);
\fill[] (Y') circle (0.8mm);
\fill[red] (Z) circle (0.8mm);
\fill[red] (Z') circle (0.8mm);
\end{scope}

\begin{scope}[shift={(9,0)}]
\defvertices
\draw[thick] (Z) (Z') -- (Y)  (Y')  (X');
\draw[thick, dotted] (Y') -- (Z);
\draw[thick, dashed] (Z') -- (Z)  (X') -- (Y');
\draw[thick, red] (Y) -- (S) -- (Y') (Z)--(X)--(R)--(X');
\draw[thick, dotted, red] (X')--(Z');
\drawvertices
\draw (0.5, -0.8) node {(3)};
\fill[] (X) circle (0.8mm);
\fill[] (X') circle (0.8mm);
\fill[red] (Y) circle (0.8mm);
\fill[red] (Y') circle (0.8mm);
\fill[red] (Z) circle (0.8mm);
\fill[red] (Z') circle (0.8mm);
\end{scope}

\begin{scope}[shift={(12,0)}]
\defvertices
\draw[thick] (Y')  (X') (Z') -- (Y)  (X) -- (Z);
\draw[thick, dotted] (Y') -- (Z) (X')--(Z');
\draw[thick, dashed] (X') -- (Y');
\draw[thick, red] (Y) -- (S) -- (Y') (X)--(R) --(X');
\draw[thick, dashed, red] (Z') -- (Z);
\drawvertices
\draw (0.5, -0.8) node {(4)};
\fill[red] (X) circle (0.8mm);
\fill[red] (X') circle (0.8mm);
\fill[red] (Y) circle (0.8mm);
\fill[red] (Y') circle (0.8mm);
\fill[red] (Z) circle (0.8mm);
\fill[red] (Z') circle (0.8mm);
\end{scope}

\end{tikzpicture}

    \caption{An example of a parity gadget (left figure). The right four figures show
    the four possibilities subtour~(1)--(4) to cover the vertices of the parity gadget by disjoint paths (red edges and red endpoints). The dashed edges are the 
    same-set edges, the dotted edges are the different-set edges, and the solid edges are the remaining edges.}
    \label{fig:parity_gadget}
\end{figure}

\begin{definition}[Parity Gadget]
\label{def:parity-gadget}
A \emph{parity gadget} is an edge weighted graph containing at least six distinct vertices labeled $X, X', Y, Y', Z, Z'$
that satisfies the following two properties.
First there exist at least the following four possibilities to cover the vertices of the parity gadget by 
vertex disjoint paths with endpoints in the set $\{X,X', Y, Y', Z, Z'\}$: 
\begin{enumerate}
    \item[(1)] A $\{Z, Z'\}$-path;
    \item[(2)] An $\{X, X'\}$-path and a $\{Z, Z'\}$-path;
    \item[(3)] A $\{Y, Y'\}$-path and a $\{Z, Z'\}$-path; or
    \item[(4)] An $\{X, X'\}$-path, a $\{Y, Y'\}$-path and a $\{Z, Z'\}$-path.
\end{enumerate}
The four possibilities are called \emph{subtour} (1), \emph{subtour} (2), \emph{subtour} (3), and \emph{subtour} (4)
(see~\cref{fig:parity_gadget} for an example). 
We require that in these four cases the cover is unique.
A parity gadget may allow more than these four possibilities
to cover the vertices by vertex disjoint paths with all endpoints in the set $\{X,X', Y, Y', Z, Z'\}$. 
Any such cover is called a \emph{subtour} as long as $Z$ and $Z'$ are endpoints of some path(s) in this cover.

We require in addition if $X$ and $X'$ are in the set of endpoints, then there must exist an $\{X,X'\}$-path in the cover. We require the same for the vertices $Y$ and $Y'$.

Second, the edges of a parity gadget must allow a partition
into three subsets, the \emph{same-set edges}, the \emph{different-set edges}, and the remaining edges.
The same-set edges have the same weight, which we call the \emph{same-set weight}.
Similarly, the different-set edges have the same \emph{different-set weight}.
The remaining edges have weight zero. 

We require that subtours (1) and (4) contain exactly one same-set edge and no different-set edge, and that subtours (2) and (3) contain exactly one different-set edge and no same-set edge. See \cref{fig:parity_gadget} for an example. 
\end{definition}

As explained before, a parity gadget is used to replace a gateway~$XX'$ of an $H$-vertex~$x$, a gateway~$YY'$ of an $H$-vertex~$y$, and the $xy$-edge~$ZZ'$.
The vertices~$X$, $X'$, $Y$, $Y'$, $Z$, and $Z'$ are part of the parity gadget, and the gadget is connected with the rest of~$G$ via exactly one incident edge to each of these vertices.
We call these six incident edges the \emph{external edges} of the gadget.
We define the \emph{internal edges} as the edges within the parity gadget.
Further, we say that the gadget is \emph{related} to the $H$-vertices $x$ and $y$.

By construction, the removed edge $ZZ'$ was originally part of a path of length five, say $(Z_1, Z_2, Z, Z', Z_3, Z_4)$.
Since $Z_2$ and $Z_3$ have degree two in~$G$, any tour of~$G$ has to contain~$Z_2Z$ and~$Z'Z_3$.
Therefore, the tour can only contain exactly one internal edge incident to~$Z$ and one incident to~$Z'$ (which may coincide). This is the reason why in the definition of a parity gadget the set $\{Z,Z'\}$ appears in
all four cases.

A subtour containing an $\{X, X'\}$-path (i.e., subtour (2) or subtour (4)) represents that the corresponding $H$-vertex $x$ is in the second set of the cut; otherwise, $x$ is in the first set.
When such a subtour occurs in the gadget, we say the gadget \emph{uses} an $\{X, X'\}$-path.
We have a similar representation and notation for the $\{Y, Y'\}$-path.

By definition of a parity gadget the total weight of the tour edges within a parity gadget is the same-set weight, when $x$ and $y$ are in the same set of the cut, and it is the different-set weight, when they are in different sets.

Next, a parity gadget is an \emph{$(r_x,r_y)$-parity gadget}, if
\begin{itemize}
    \item We need to remove exactly~$r_x$ internal edges and add exactly~$r_x-1$ internal edges to change from subtour (1) to subtour (2) or from subtour (3) to subtour (4); 
    \item We need to remove exactly~$r_y$ internal edges and add exactly~$r_y-1$ internal edges to change from subtour (1) to subtour (3) or from subtour (2) to subtour (4); 
    \item In order to change from subtour (1) to subtour (4) or between subtour (2) and subtour (3), we need to remove at least $\max\{r_x, r_y\}$ internal edges and add at least $\max\{r_x, r_y\}-1$ internal edges.
\end{itemize}

We call the changes in the first two conditions above and their reverses the \emph{standard subtour changes}.

By definition, a parity gadget may allow more than the four subtours~(1)--(4) as possibilities to cover the vertices
by disjoint paths with end vertices in the set $\{X, X', Y, Y', Z, Z'\}$. The parity gadget shown in \cref{fig:parity_gadget}
allows for example to cover the vertices by a $\{Z,X'\}$-path and a $\{Z',Y'\}$-path. 
We say that a parity gadget is a \emph{strict parity gadget}, or simply a \emph{strict gadget}, if subtours (1)-(4) are the only possible subtours for the gadget in $G$, \emph{after} we equip all gadgets (details are given in \cref{sec:equip}) used in the reduction. Thus the property of being strict may depend on the other gadgets used in the reduction.

\subsection{XOR gadgets}
\label{sec:xor_gadgets}
\begin{figure}
    \centering
    \includegraphics{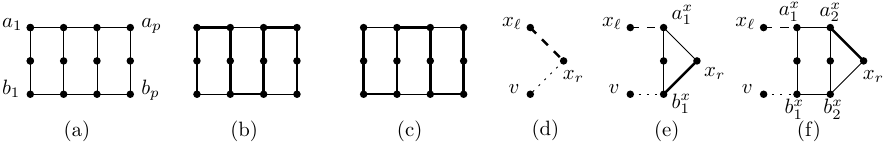}
    \caption{The XOR gadget of order four (a) and its two subtours ((b) and (c)). 
    (d)-(f) are examples of assigning the XOR gadgets of order from zero to two to the $H$-vertex~$x$, where dashed edges, dotted edges, and bold edges represent the left first-set edge, the door closest to $x_r$, and the right first-set edge, respectively.}
    \label{fig:xor_gadget}
\end{figure}

The remaining gadgets used in the reduction are the XOR gadgets.
We generalize these gadgets from the XOR gadget by~\cite{pap1978}.
See \cref{fig:xor_gadget}(a)-(c) for an illustration for the definition below.

\begin{definition}[XOR Gadget]
\label{def:xor_gadget}
    Let $p$ be a nonnegative integer.
    The \emph{XOR gadget} of order~$p$ is a graph containing two paths $(a_1, \dots, a_{p})$ and $(b_1, \dots, b_{p})$, and for $i \in \{1, \dots, p\}$, there is a path of length two with $a_i$ and $b_i$ as endpoints.
    A \emph{subtour} of the XOR gadget is a spanning path with two endpoints in the set $\{a_1, a_p, b_1, b_p\}$.
    For convenience, when $p = 0$, both the XOR gadget of order zero and its only subtour are defined to be the empty graph.
\end{definition}

It is easy to see that an XOR gadget has two subtours, except when $p \leq 1$, in which case, it has only one subtour.
Further, for $p \geq 2$, changing from one subtour to the other requires a swap of $p-1$ edges.

We define the \emph{assigning} of the XOR gadget of some order~$p$ to an $H$-vertex~$x$ as follows. 
(See~\cref{fig:xor_gadget}(d)-(f) for an illustration.)
Recall that $x_{\ell}x_r$ is the first-set edge of $x$, and let $vx_r$ be the incident door of~$x$ to $x_r$.
We subdivide the two edges above into paths of length~$p+1$, $(x_{\ell}, a^x_1, \dots, a^x_p, x_r)$ and $(v, b^x_1, \dots, b^x_p, x_r)$.
Then for $i \in \{1, \dots, p\}$, we connect $a^x_i$ and $b^x_i$ with a path of length two.
Note that when $p = 0$, we do nothing.
Further note that when we remove the edges incident to $x_{\ell}$, $x_r$, and $v$ in the construction above, we obtain the XOR gadget of order~$p$ as defined in \cref{def:xor_gadget}.

We call these incident edges to $x_{\ell}$, $x_r$, and $v$ the \emph{external edges} of the XOR gadget, and we call the other edges in the construction the \emph{internal edges} of the gadget.
For convenience, we still refer to the external edge incident to $v$ (i.e., $vb^x_1$ for $p \geq 1$ and $vx_r$ for $p=0$) as a door of $x$.
Additionally, we call it the \emph{closest door to $x_r$}.
We call the external edge incident to $x_{\ell}$ (i.e., $x_{\ell}a^x_1$ for $p \geq 1$ and $x_{\ell}x_r$ for $p = 0$) the \emph{left first-set edge} of~$x$. 
We define the \emph{right first-set edge} of~$x$ as $x_rx_{\ell}$ if $p = 0$, $x_ra^x_p$ if $p$ is positive and even, and $x_rb^x_p$ if $p$ is odd.
Lastly, we call the other external edge incident to $x_r$ the \emph{right second-set edge} of~$x$.

We define an \emph{incident edge} of a nonempty subtour of the XOR gadget to be an external edge incident to an endpoint of the subtour.
The incident edge of an empty subtour (i.e., when $p = 0$) is defined to be simply an external edge of the XOR gadget.

Based on the definitions above, it is easy to verify the following.
\begin{observation}
\label{obs:xor_external_edges}
    For the XOR gadget assigned to an $H$-vertex~$x$, one subtour of the gadget is incident to the left and right first-set edges of~$x$, and another subtour of the gadget is incident to the closest door to~$x_r$ and the right second-set edge of~$x$.
    The two subtours above are identical, if the order of the gadget is at most one.
    Otherwise, they are distinct.
\end{observation}

\subsection{Equipping gadgets}
\label{sec:equip}
Within our reduction we will use different parity gadgets at different places and XOR gadgets of different orders. 
The exact specification of which parity gadget we use at what place and the XOR gadget of which order is used for which $H$-vertex will be called the \emph{equipping} of gadgets.
We equip the gadgets through a labeling scheme.
Specifically, for an $H$-vertex $x$ and an incident $H$-edge $z$, a labeling~$L$ assigns an integer label to the pair $(x,z)$.
We say the label is incident to $x$ and to $z$.
Additionally, $L$ also assigns an integer label to each $H$-vertex.
We also say this label is incident to $x$.
We call a labeling~$L$ \emph{valid}, if for every $H$-edge~$xy$, there exists a $(L(x,xy),L(y,xy))$-parity gadget, and for every $H$-vertex~$x$, $L(x) \geq 0$.

Then we equip the gadgets based on a valid labeling~$L$ as follows.
For an $H$-edge $xy$, we equip the $(L(x,xy),L(y,xy))$-parity gadget to $xy$.
Recall that $w(xy)$ is the weight of $xy$.
If $w(xy) \geq 0$, the same-set weight of the gadget is $w(xy)$, and its different-set weight is zero.
Otherwise, the same-set weight of the gadget is zero, and its different-set weight is $-w(xy)$.
Finally, for an $H$-vertex~$x$, we assign the XOR gadget of order~$L(x)$ to $x$, and all edges involved in this assignment have weight zero.
We call this a \emph{gadget arrangement} corresponding to $L$. 
Note that this construction implies that edge weights in the TSP instance are nonnegative. 

For each $H$-vertex $x$, we define the \emph{label sum} of $x$ to be the sum of the labels incident to $x$, i.e., $L(x) + \sum_{y:xy\in E(H)} L(x, xy)$.
We say a labeling is an $s$-labeling, if each $H$-vertex has label sum $s$. 
By the gadget arrangement above, we have a simple lemma:
\begin{lemma}
\label{obs:x_change_12}
    For an $H$-vertex $x$, let $d(x)$ be its degree in $H$ and $s$ be its label sum with respect to some valid labeling~$L$.
    Suppose that in the corresponding gadget arrangement, the parity gadgets related to~$x$ use either subtour~(1) or subtour~(3) (i.e., none of them uses an $\{X,X'\}$-path).
    Then in order to change the subtours in all these parity gadgets to either subtour~(2) or subtour~(4) (i.e., after the change, they all use an $\{X,X'\}$-path), we need to remove at least $s - L(x)$ internal edges and add at least $s - L(x) - d(x)$ internal edges across all these parity gadgets.
    Further, we remove and add exactly the numbers stated above, if we only use standard subtour changes.
\end{lemma}
\begin{proof}
    We first assume that we only use the standard subtour changes (i.e., changes from subtour~(1) to subtour~(2) or from subtour~(3) to subtour~(4)).
    Then by the definition of an $(r_x, r_y)$-parity gadget and the label sum, we need to remove exactly $s - L(x)$ internal edges and add exactly $s - L(x) - d(x)$ internal edges.

    Now, suppose our assumption above is not true; that is, we change either from subtour~(1) to subtour~(4) or from subtour~(2) to subtour~(3), for some parity gadgets. 
    Then by the last condition in the definition of an $(r_x, r_y)$-parity gadget, we must remove and add at least as many edges as we need for the standard subtour changes.

    The lemma then follows.
\end{proof}

\subsection{Initial tour}
To complete the description of the TSP/\kopt instance, we specify the initial tour of $G$.
We obtain this tour from the initial cut of $H$ as follows.
The tour contains all incident edges of degree-two vertices.
For an $H$-vertex $x$, if $x$ is in the first set, we include the left and right first-set edges of $x$ in the tour.
Further, we use the subtour of the XOR gadget assigned to $x$, such that this subtour is incident to the left and right first-set edges of~$x$.
If $x$ is in the second set, we include all the doors and the right second-set edge of $x$ in the tour.
Moreover, we use the subtour of the XOR gadget assigned to $x$, such that this subtour is incident to the right second-set edge of~$x$ and the closest door to $x_r$.

For an $H$-edge $xy$, in the corresponding gadget, we use the subtour (1) if $x$ and $y$ are in the first set, subtour (2) if $x$ is in the second set and $y$ in the first set, subtour (3) if $x$ is in the first set and $y$ in the second set, and subtour (4) if $x$ and $y$ are in the second set.

By the construction of $G$ and the definition of the subtours, it can be verified that we obtain a tour of $G$.

\subsection{Correspondence between flip sequences and swap sequences}
\label{sec:correspondence}
For the rest of this section, we assume that we have a valid $s$-labeling $L$, and the parity gadgets in the corresponding gadget arrangement are strict.

We start with a simple lemma about doors of an $H$-vertex.

\begin{lemma}
\label{lem:end_doors}
    For any tour of $G$ and any $H$-vertex~$x$, the tour contains either both the door incident to $x_{\ell}$ and the closest door to ${x_r}$ or none of them.
    Further, these two doors are in the tour, if and only if the left and right first-set edges of $x$ are not in the tour.
\end{lemma}
\begin{proof}
    By construction, $x_{\ell}$ is incident to three edges: the left first-set edge, a door, and an edge incident to a degree-two vertex.
    Since the last edge has to be in the tour, the tour uses either the left first-set edge or the door incident to $x$.
    By a similar argument, the tour uses either the right first-set edge or the right second-set edge of $x$.
    Lastly, the tour has to use a subtour of the XOR gadget assigned to~$x$, if the order of the XOR gadget is nonzero.
    Combined with \cref{obs:xor_external_edges}, all of the above implies the lemma.
\end{proof}

Next, we have the following lemma about the edges in a tour of $G$.

\begin{lemma}
\label{lem:tour_edges}
    For any tour of $G$ and for any $H$-vertex~$x$, exactly one of these cases holds:
    \begin{itemize}
        \item The first-set case: The tour uses the left and right first-set edges of~$x$ and does not use any doors or the right second-set edge of~$x$. 
        It also uses the subtour of the XOR gadget assigned to~$x$ incident to the left and right first-set edges.
        Further, in all parity gadgets related to $x$ the tour does not use an $\{X,X'\}$-path.
        \item The second-set case: The tour uses all the doors and the right second-set edge of~$x$ and does not use the left and right first-set edges of~$x$. 
        It also uses the subtour of the XOR gadget assigned to~$x$ incident to the right second-set edge of $x$ and the closest door to~$x_r$.
        Further, in all parity gadgets related to $x$ the tour uses an $\{X,X'\}$-path.
    \end{itemize}
\end{lemma}
\begin{proof}
    Recall that the second-set path of $x$ was $(x_{\ell}, x_{1}, x'_{1}, \dots, x_{d(x)}, x'_{d(x)}, x_r)$, where $d(x)$ denotes the degree of $x$ in $H$.
    For ease of argument, we define $x'_0 := x_{\ell}$, and we define $x_{d(x)+1}$ to be the endpoint of the closest door to $x_r$ other than $x_{d(x)}$ (i.e., $x_{d(x)}x_{d(x)+1}$ is an external edge of the XOR gadget assigned to~$x$.
    
    Firstly, we show that either all the doors of $x$ are in the tour or all doors of $x$ are not in the tour.
    For the sake of contradiction, suppose that this does not hold.
    This implies that there are two edges $x'_{i-1} x_i$ and $x'_i x_{i+1}$ such that one of them is in the tour and the other one is not.
    Without loss of generality, assume that $x'_{i-1} x_i$ is in the tour.
    Note that other than these two edges, $x_i$ and $x'_i$ are only incident to internal edges of the parity gadget that replaces the gateway $x_i x'_i$.
    All of the above imply that the parity gadget must use a path that starts at $x_i$ and ends at another vertex that is not $x'_i$.
    However, since the parity gadget is strict, and none of the subtours (1)-(4) allow such a path, we have a contradiction.
    
    Lastly, the above fact implies that when all doors are in the tour, the gadgets related to $x$ have to use an $\{X,X'\}$-path.
    Likewise, when all these doors are not in the tour, the gadgets must not use an $\{X,X'\}$-path.
    
    The lemma then follows from the two preceding paragraphs, \cref{obs:xor_external_edges}, \cref{lem:end_doors}, and the fact that the tour has to use a subtour of the XOR gadget assigned to~$x$.
\end{proof}

For an $H$-vertex $x$, we define an \emph{$x$-change} in a tour as the swap that can be performed to change from the first-set case of \cref{lem:tour_edges} to the second-set case, or vice versa. 
The following lemma shows that such a swap always exists and 
it specifies the number of edges to be swapped for an $x$-change.

\begin{lemma}
\label{lem:x-change}
    For an $H$-vertex $x$, an $x$-change is a swap of at least $s+1$ edges.
    Further, if we use only standard subtour changes, then it is a swap of exactly $s+1$ edges.
\end{lemma}
\begin{proof}
    We consider the $x$-change from the first-set case of \cref{lem:tour_edges} to the second-set case.
    For the other direction, we just switch the two sets of the swap, and hence the number of edges involved is the same.

    Let $d(x)$ be the degree of $x$ in $H$.
    There are three parts of the $x$-change.
    Firstly, we need to make the change in every parity gadget related to $x$ so that it uses an $\{X, X'\}$-path (i.e., it uses either subtour~(2) or subtour~(4)).
    By \cref{obs:x_change_12}, we can do this by removing at least $s - L(x)$ edges and add at least $s - L(x) - d(x)$ edges.
    Moreover, if we use only standard subtour changes, we remove and add exactly these numbers of edges. 
    Secondly, we need to remove the left first-set edge and add $d(x)+1$ doors.
    Lastly, if the order of the XOR gadget assigned to $x$ (i.e., $L(x)$) is nonzero, we need to remove the right first-set edge and add the right second-set edge.
    In addition, we need to change the subtour of the XOR gadget, by removing and adding $L(x) - 1$ edges.
    In any case, we add at least $s+1$ edges and remove at least $s+1$ edges in total, and the numbers are exact when we use only standard subtour changes.
    The lemma then follows.
\end{proof}

Based on the construction of the initial tour, the pair of the initial cut and the initial tour satisfies the following property (*): For an $H$-vertex $x$, $x$ is in the first set, if and only if the first-set case of \cref{lem:tour_edges} holds for $G$ and $x$.

We now prove the correspondence between the flipping sequence in $H$ and the $(s+1)$-swap sequence in $G$ in the following two lemmata.

\begin{lemma}
\label{lem:correspondence_13_1}
    Suppose there are a cut $\gamma_1$ and a tour $\tau_1$, such that $(\gamma_1, \tau_1)$ has property (*).    
    Then for any cut $\gamma_2$ obtained from $\gamma_1$ by an improving flip, there exists a tour $\tau_2$, obtained from $\tau_1$ by an improving $(s+1)$-swap, such that $(\gamma_2, \tau_2)$ also has property (*).
\end{lemma}
\begin{proof}
    Suppose the cuts $\gamma_1$ and $\gamma_2$ differ by an $H$-vertex $x$, where $x$ is in different sets in $\gamma_1$ and $\gamma_2$.
    Let $\tau_2$ be the tour obtained by performing an $x$-change on $\tau_1$ with only standard subtour changes.
    By the definition of $x$-change, it is easy to see that $(\gamma_2, \tau_2)$ has property (*).
    Further, by \cref{lem:x-change}, we can transform $\tau_1$ to $\tau_2$ by performing an $(s+1)$-swap.

    Hence, what remains to be shown is that the $x$-change is improving.
    Suppose $P$ and $Q$ are the two sets of $\gamma_1$, such that $x \in P$. (We do not assume which one of $P$ and $Q$ is the first set.) 
    Then for the change from $\gamma_1$ to $\gamma_2$, the change in value for the Max-Cut instance is
    \[
      - \sum_{xy \in E(H) : y \in Q} w(xy) + \sum_{xy \in E(H) : y \in P} w(xy) := \Delta.
    \]
    Since $\gamma_2$ is the result of an improving flip from $\gamma_1$, we have $\Delta > 0$.

    Recall that the only tour edges with non-zero weights are the same-set and different-set edges in the gadgets, and the total edge weight of a gadget is the same-set weight or the different-set weight, depending on whether the presence statuses of an $\{X,X'\}$-path and a $\{Y, Y'\}$-path are the same or different. 
    Due to the property (*) of $(\gamma_1, \tau_1)$ and $(\gamma_2, \tau_2)$ and due to the assignment of the same-set and different-set weights to the gadgets, the change of the total edge weight of the tour from $\tau_1$ to $\tau_2$ is
    \begin{align*}
        & - \Big(\sum_{\substack{xy \in E(H) : \\ y \in P, w(xy) \geq 0}} w(xy) - \sum_{\substack{xy \in E(H) : \\ y \in Q, w(xy) < 0}} w(xy) \Big) + \Big(\sum_{\substack{xy \in E(H) : \\ y \in Q, w(xy) \geq 0}} w(xy) - \sum_{\substack{xy \in E(H) : \\ y \in P, w(xy) < 0}} w(xy) \Big) \\
        = & \sum_{\substack{xy \in E(H) : \\ y \in Q}} w(xy) - \sum_{\substack{xy \in E(H) : \\ y \in P}} w(xy) = -\Delta < 0.
    \end{align*}

    The lemma then follows.    
\end{proof}

The next lemma shows the reverse correspondence.
\begin{lemma}
\label{lem:correspondence_13_2}
    Suppose there are a cut $\gamma_1$ and a tour $\tau_1$, such that the pair $(\gamma_1, \tau_1)$ has property (*).
    Then for any tour $\tau_2$, obtained from $\tau_1$ by performing an improving $(s+1)$-swap, there exists a cut $\gamma_2$ obtained from $\gamma_1$ by an improving flip, such that $(\gamma_2, \tau_2)$ also has property (*).
\end{lemma}
\begin{proof}
    If $\tau_2$ is obtained from $\tau_1$ by performing an $x$-change for some $H$-vertex $x$, then we can define~$\gamma_2$ to be the cut obtained from~$\gamma_1$ by flipping $x$.
    Using an analogous argument as in the proof of \cref{lem:correspondence_13_1}, we can see that $\gamma_2$ satisfies the lemma's statement.
    Hence, it remains to prove that any $(s+1)$-swap has to be an $x$-change, for some $x$.

    By \cref{lem:tour_edges}, if we remove a first-set edge of an $H$-vertex~$x$, a door of $x$, or the second-set edge of~$x$, then we have to do an $x$-change in order to maintain a tour.
    Since the $x$-change already involves a swap of at least $s+1$ edges, we cannot swap any other edge.
    If we do not change any first-set edge or second-set path, then by the design of parity gadgets, we cannot change the subtour of any gadget.
    As a result, the tour remains the same.
    This completes the proof of the lemma.
\end{proof}

Lemmata~\ref{lem:correspondence_13_1} and~\ref{lem:correspondence_13_2} establish a one-to-one correspondence between improving flip sequences and improving $(s+1)$-swap sequences, as follows.

\begin{lemma}
\label{lem:correspondence}
    Let $I$ be a Max-Cut/Flip instance and $\sigma$ be an initial cut for $I$.
    Suppose for some $s$, there is a valid $s$-labeling~$L$ for $I$ such that all gadgets in the gadget arrangement in $L$ are strict.
    Then we can reduce $I$ to a TSP/\kopt instance~$I'$, for $k = s+1$, and obtain an initial tour~$\tau$ from~$\sigma$, such that there is a one-to-one correspondence between improving flip sequences for $I$ and $\sigma$ and improving $k$-swap sequences for $I'$ and $\tau$.
\end{lemma}

\section{All-exp property for $k \geq 13$}
\label{sec:all-exp-13}
In this section, we prove the following statement. 

\begin{lemma}
\label{lem:all_exp_13}
    TSP/\kopt has the all-exp property for $k \geq 13$.
\end{lemma}

\subsection{Simple gadgets}
We start with the parity gadgets.
We call a parity gadget a \emph{simple gadget}, if the parity gadget only allows subtours (1)--(4) (and no other subtours), independent of other gadgets equipped in the graph $G$. If a simple gadget is an $(r_x,r_y)$-parity gadget we call it an $(r_x,r_y)$-simple gadget. 
Note that a simple gadget is a strict gadget.

In this section, we provide a $(4,2)$-simple gadget (and by symmetry, we also have a $(2,4)$-simple gadget).

\begin{figure}[t!]
    \centering
    \includegraphics[scale=0.75]{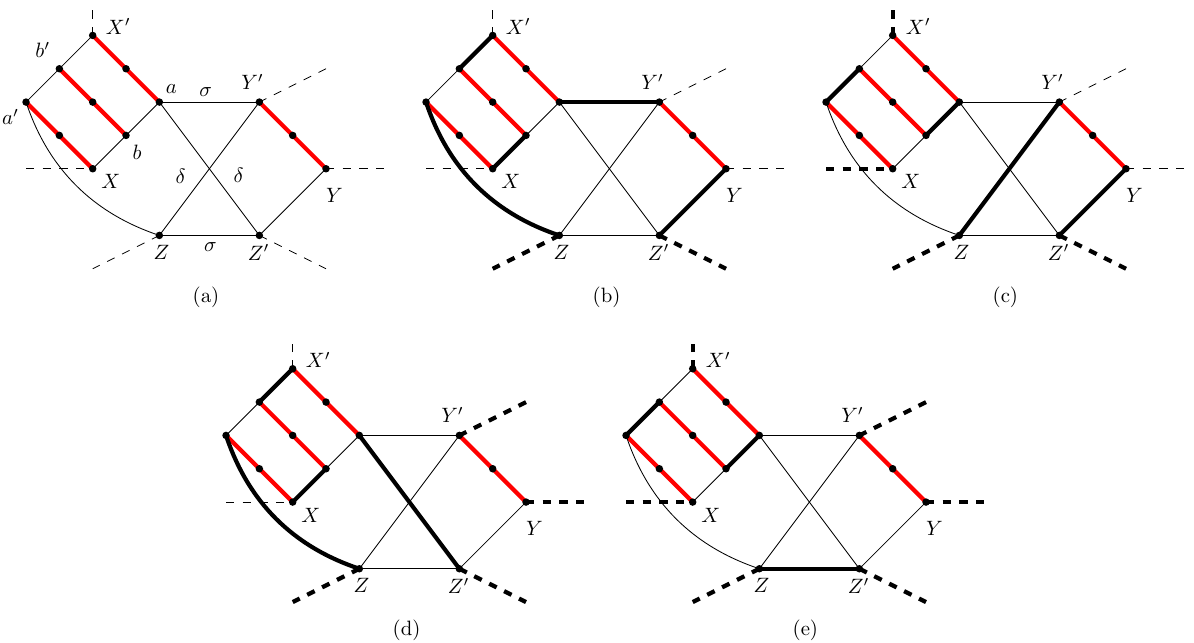}
    \caption{(a) A (4,2)-simple gadget. Bold red edges are incident to degree-two vertices. Dashed edges are external edges. $Y'a$ and $ZZ'$ are same-set edges with same-set weight $\sigma$, while $Z'a$ and $Y'Z$ are different-set edges with different-set weight $\delta$. (b)-(e) show the subtours (1)-(4). }
    \label{fig:strict_gadget_1}
\end{figure}

\begin{lemma}
\label{lem:simple_gadgets}
    The gadget as depicted in \cref{fig:strict_gadget_1} is a $(4,2)$-simple gadget.
\end{lemma}
\begin{proof}

    Let $g$ be the gadget as depicted in \cref{fig:strict_gadget_1}.
    Let $g_x$ be the XOR gadget containing nine vertices including $X, X', a, a'$, such that its two subtours are an $\{X, X'\}$- and an $\{a, a'\}$-path.
    We can assume that the two edges of the path of length two between $Y$ and $Y'$ are always in a path cover of~$g$.
    
    Firstly, suppose that $X$ is an endpoint of a path in the path cover~$g$.
    Then we know that the XOR gadget $g_x$ uses the $\{X,X'\}$ path.
    This implies that the edges $Za'$, $Z'a$, and $aY'$ are not in a path from the path cover.
    By definition of a parity gadget the vertices $Z$ and $Z'$ must be endpoints of paths in the path cover.  
    If $ZZ'$ is an edge in the path cover, then we get the subtour~(4).
    If  $ZZ'$ is not an edge then $ZY'$ and $Z'Y$ must be edges in the path cover. Thus we get subtour~(2).

    Secondly, assume that $X$ is not an endpoint of a path in the path cover of the gadget.
    Then the XOR gadget $g_x$ implies that we have to use the edge $Za'$ in the path cover and $X'$ cannot be the endpoint of a path in the path cover. 
    If $Z'a$ is an edge in the path cover then we get subtour~(3). 
    If this is not the case we get subtour~(1).

    The two preceding paragraphs imply that the only possible subtours in $g$ are the four depicted in \cref{fig:strict_gadget_1}.
    From the figure, it can be checked that the requirements regarding the same-set and different-set edges are met.
    Hence, $g$ is a simple gadget.

    Lastly, it can also be checked that the gadget is a $(4,2)$-simple gadget.
\end{proof}

One can show that the above result is tight in the sense that there are no $(2,2)$- or $(2,3)$-simple gadgets.

\subsection{Proof of \cref{lem:all_exp_13}}
We use the reduction from Max-Cut/Flip to TSP/\kopt as described in \cref{sec:reduction}.
By \cref{thm:all_exp_maxcut} we may assume $H$ to have maximum degree four.

We now construct a valid $(k-1)$-labeling.
Firstly, we assign an orientation on the $H$-edges, such that every degree-four $H$-vertex has exactly two incoming edges and two outgoing edges.
Specifically, we repeat the following procedure: Until all $H$-edges have an orientation, we take a maximal (possibly closed) walk in the subgraph of unoriented $H$-edges, and we orient the edges along the walk.
It is easy to see that the orientation is as desired.

Next, for every directed $H$-edge $z$ with head $x$ and tail $y$, we label $(x,z)$ and $(y,z)$ with four and two, respectively.
By \cref{lem:simple_gadgets}, there exists a simple gadget corresponding to these labels.
Our construction guarantees that up to this point the label sum of every $H$-vertex is at most~12. 
Next, for each $H$-vertex, we assign a nonnegative label to the $H$-vertex, such that the label sum at the $H$-vertex is exactly $k-1$. This is possible because $k \geq 13$.
Hence, we obtain a valid $(k-1)$-labeling.

Further, since we use only simple gadgets, the labeling above satisfies the condition of \cref{lem:correspondence}. 
Then by \cref{thm:all_exp_maxcut} and \cref{lem:correspondence}, we obtain \cref{lem:all_exp_13}.

\section{All-exp property for $k \geq 9$}
In this section, we prove the following statement. 

\begin{lemma}
\label{lem:all_exp_9}
    TSP/\kopt has the all-exp property for $k \geq 9$.
\end{lemma}

We use the reduction in \cref{sec:reduction}, with two extra ingredients.
Firstly, we look into the construction by Michel and Scott~\cite{Michel_Max_Cut_4}.
Secondly, we introduce a (2,2)-parity gadget called the flexible gadget.
We discuss these two aspects in more detail in Sections~\ref{sec:michel_scott} and~\ref{sec:flexible}.
After that, we combine them with the reduction in \cref{sec:reduction} to obtain the proof of \cref{lem:all_exp_9}.

\subsection{Michel-Scott construction of Max-Cut/Flip instance}
\label{sec:michel_scott}

This subsection describes the construction of the Max-Cut/Flip instance by Michel and Scott~\cite{Michel_Max_Cut_4}.

\begin{figure}[t!]
    \centering
    \includegraphics[page=1]{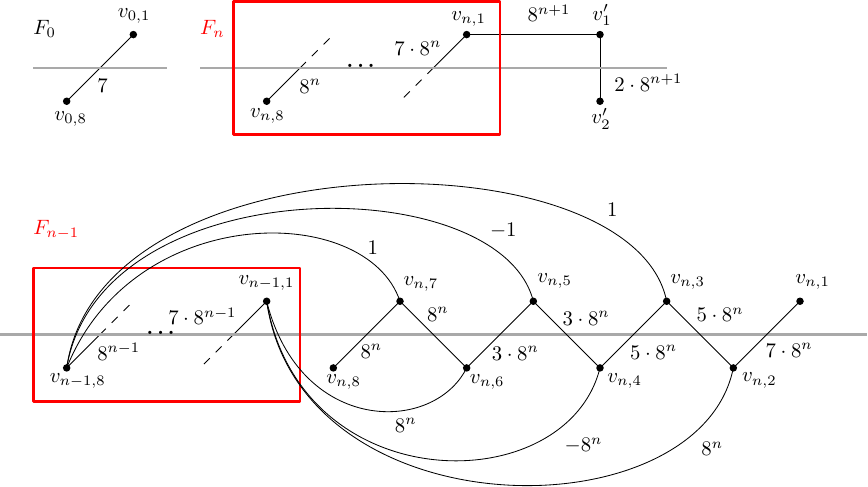}
    \caption{Michel-Scott construction of a Max-Cut instance that has an exponentially long improving flip sequence. Vertices on one side of the horizontal line are in the same set of the initial cut. Figure adopted from \cite[Figure 1]{Michel_Max_Cut_4}.}
    \label{fig:michel_scott_1}
\end{figure}

The graph for the instance is constructed inductively as follows (see \cref{fig:michel_scott_1} for an illustration):
\begin{itemize}
    \item The base graph $F_0$ consists of a single edge $v_{0,1} v_{0,8}$ with weight 7.
    \item The graph $F_n$ contains a path of eight new vertices $v_{n,1}, v_{n,2}, v_{n,3}, v_{n,4}, v_{n,5}, \allowbreak v_{n,6}, v_{n,7}, \allowbreak v_{n,8}$ that appear in the path in that order.
    The weights of the edges along the path from $v_{n,1}$ to $v_{n,8}$ are 
    $7 \cdot 8^n,  5 \cdot 8^n, 5 \cdot 8^n, 3 \cdot 8^n, 3 \cdot 8^n, 8^n, 8^n$.
    Next, we connect $F_n$ to $F_{n-1}$ as follows: We add edges connecting $v_{n-1,1}$ to $v_{n,2}, v_{n,4}$, and $v_{n,6}$, with weights $8^n, -8^n$, and $8^n$, respectively.
    Finally, we add edges connecting $v_{n-1,8}$ to $v_{n,3}, v_{n,5}$, and $v_{n,7}$, with weights 1, -1, and 1, respectively.
\end{itemize}

The final graph~$H_n$ consists of all the graphs $F_0, \dots, F_n$ and the connecting edges, as well as two new vertices $v'_1$ and $v'_2$ and two new edges $v_{n,1} v'_1$ and $v'_1 v'_2$ with weights $8^{n+1}$ and $2 \cdot 8^{n+1}$, respectively.
The initial cut of $H_n$ is as follows:
One set of the cut contains exactly $v'_1$ and all vertices $v_{i,j}$ such that $j$ is odd.
The other set contains the remaining vertices.

Michel and Scott~\cite{Michel_Max_Cut_4} showed that there exists a unique maximal improving flip sequence~$L_n$ from the aforementioned cut of $H_n$.
The sequence is described recursively as follows:
$L_0 = v_{0,1} v_{0,8}$, and 
\[
L_n = v_{n,1} v_{n,2} L_{n-1} v_{n,3} v_{n,4} L_{n-1} v_{n,5} v_{n,6} L_{n-1} v_{n,7} v_{n,8}.
\]

\subsection{Flexible gadget and forcing rule}
\label{sec:flexible}

\begin{figure}[ht]
    \centering

\begin{tikzpicture}[scale=1.0]
\small
\def\defvertices{
\coordinate[label=left: $X$] (X) at (0,3);
\coordinate[label=left: $Y$] (Y) at (0,0);
\coordinate[label=left: $Z$] (Z) at (0,2);
\coordinate[label=right: $X'$] (X') at (1,2);
\coordinate[label=right: $Y'$] (Y') at (1,1);
\coordinate[label=left: $Z'$] (Z') at (0,1);
\coordinate[] (R) at (1,3);
\coordinate[] (S) at (1,0);
}
\def\drawvertices{
\fill[] (R) circle (0.8mm);
\fill[] (S) circle (0.8mm);}

\defvertices
\draw[thick, red] (Y') -- (S) -- (Y) -- (Z') (Z) -- (X) -- (R) -- (X');
\draw[thick, dashed] (Z') -- (Z)  (X') -- (Y');
\draw[thick, dotted] (Y') -- (Z) (X')--(Z');
\drawvertices
\fill[] (X) circle (0.8mm);
\fill[red] (X') circle (0.8mm);
\fill[] (Y) circle (0.8mm);
\fill[red] (Y') circle (0.8mm);
\fill[red] (Z) circle (0.8mm);
\fill[red] (Z') circle (0.8mm);

\begin{scope}[shift={(3,0)}]
\defvertices
\draw[thick] (Y')  (S)  (Y) (Z') (Z) (X) (R) (X')  (Z) -- (X)  (Y)--(Z');
\draw[thick, dotted, red] (Y') -- (Z) (X')--(Z');
\draw[thick, dashed] (Z') -- (Z)   (X') -- (Y');
\draw[thick, red] (X)--(R)--(X') (Y')--(S)--(Y);
\drawvertices
\fill[red] (X) circle (0.8mm);
\fill[] (X') circle (0.8mm);
\fill[red] (Y) circle (0.8mm);
\fill[] (Y') circle (0.8mm);
\fill[red] (Z) circle (0.8mm);
\fill[red] (Z') circle (0.8mm);
\end{scope}

\begin{scope}[shift={(6,0)}]
\defvertices
\draw[thick] (X) -- (Z) (Z')  (Y') (X') (Z') -- (Y);
\draw[thick, dotted] (X')--(Z') (Y') -- (Z);
\draw[thick, dashed, red] (Z') -- (Z) (X') -- (Y');
\draw[thick, red] (Y')--(S)--(Y) (X)--(R)--(X');
\drawvertices
\fill[red] (X) circle (0.8mm);
\fill[] (X') circle (0.8mm);
\fill[red] (Y) circle (0.8mm);
\fill[] (Y') circle (0.8mm);
\fill[red] (Z) circle (0.8mm);
\fill[red] (Z') circle (0.8mm);
\end{scope}

\end{tikzpicture}

\caption{The subtours of the flexible gadget other than subtours (1)-(4).}
\label{fig:flexible_gadget}
\end{figure}
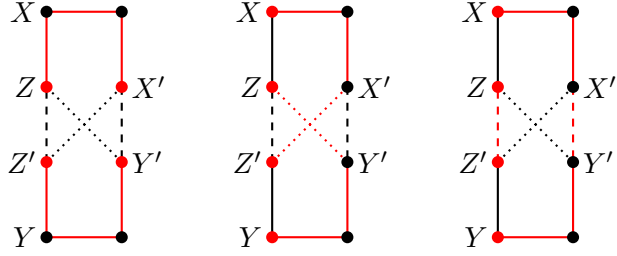

In this section, we discuss the $(2,2)$-parity gadget as depicted in \cref{fig:parity_gadget}.
We call this gadget the \emph{flexible gadget}.
Note that it is not a simple gadget, because besides the subtours (1)-(4), there are other subtours, as shown in \cref{fig:flexible_gadget}.
However, we show that a suitable equipping of the flexible and simple gadgets can force the flexible gadgets to use only the subtours (1)-(4), and hence, they are strict.

In order to describe the forcing rules, we first have some definitions related to strictness.
Note that for all the following definitions, we assume that $G$ is fixed (i.e., we have equipped all gadgets).
Suppose we have a parity gadget related to an $H$-vertex $x$.
We say that the gadget is \emph{strict at $x$}, if any tour in $G$ has to use either both external edges that are also doors of $x$ or none of these two edges.
In other words, if we look at the subtour restricted to the internal edges of the gadget, the degrees of $X$ and $X'$ have to be the same.
Put differently, the subtour has either no paths with $X$ or $X'$ as endpoints or a path with $X$ as an endpoint and a (not necessarily different) path with $X'$ as an endpoint.

Observe that for the flexible gadget, the subtours other than subtours~(1)-(4) have exactly one of $X$ and $X'$ as endpoint of a path and the other one not.
Hence, if the flexible gadget is strict at both of its related $H$-vertices, then it is strict.
Note that simple gadgets are always strict at their related $H$-vertices.

Next, the following lemmata describe two forcing rules of strictness.

\begin{lemma}
\label{lem:forcing_a}
    If a parity gadget is strict at one related $H$-vertex, then it is also strict at the other related $H$-vertex.
\end{lemma}
\begin{proof}
    Consider a parity gadget that replaces the gateways of $x$ and $y$ and the $xy$-edge.
    Further suppose that the gadget is strict at $x$.
    That means that either both $X$ and $X'$ or none of them are endpoints of paths in a subtour of the gadget.
    By the definition of a parity gadget, $Z$ and $Z'$ are endpoints of paths in any subtour.
    Because the number of endpoints is even, we obtain strictness at $y$, which is what is required to be shown.
\end{proof}

\begin{lemma}
\label{lem:forcing_b}
    Suppose $x$ is an $H$-vertex with degree $d(x)$.
    If $d(x)-1$ parity gadgets related to $x$ are strict at $x$, then the remaining parity gadget related to $x$ is also strict at $x$.
\end{lemma}
\begin{proof}
    By \cref{lem:end_doors}, the door incident to $x_{\ell}$ and the closest door to $x_r$ have to be either both present or both absent in the tour.
    Together with the strictness of the other $d(x)-1$ gadgets, this implies that either all doors of $x$ are in the tour or all doors of $x$ are not in the tour.
    This implies the strictness at $x$ for the remaining gadget.
\end{proof}

From the forcing rules above, we have a sufficient condition of the equipping of simple and flexible gadgets to ensure that all parity gadgets are strict.

\begin{lemma}
\label{lem:strict_flexible_gadgets}
    Suppose in the reduction in \cref{sec:reduction}, we only equip simple, flexible, and XOR gadgets.
    If the $H$-edges corresponding to the flexible gadgets form a forest in $H$, then all parity gadgets are strict.
\end{lemma}
\begin{proof}
    The lemma is obtained from a simple recursive argument on a leaf of the forest, as follows.
    Consider the subgraph of $H$ containing all $H$-edges corresponding to the flexible gadgets that have not been proved to be strict.
    Then the subgraph is a forest.
    Applying \cref{lem:forcing_b} to a leaf of the forest, we obtain the strictness at the leaf $H$-vertex for the flexible gadget corresponding to the edge incident to this leaf.
    Combined with \cref{lem:forcing_a}, this flexible gadget is strict at both its related $H$-vertices.
    This implies that it can only use subtours~(1)-(4), and hence, it is strict. 
\end{proof}

\subsection{Proof of \cref{lem:all_exp_9}}
\label{sec:proof_all_exp_9}
\begin{figure}
    \centering
    \includegraphics[page=2]{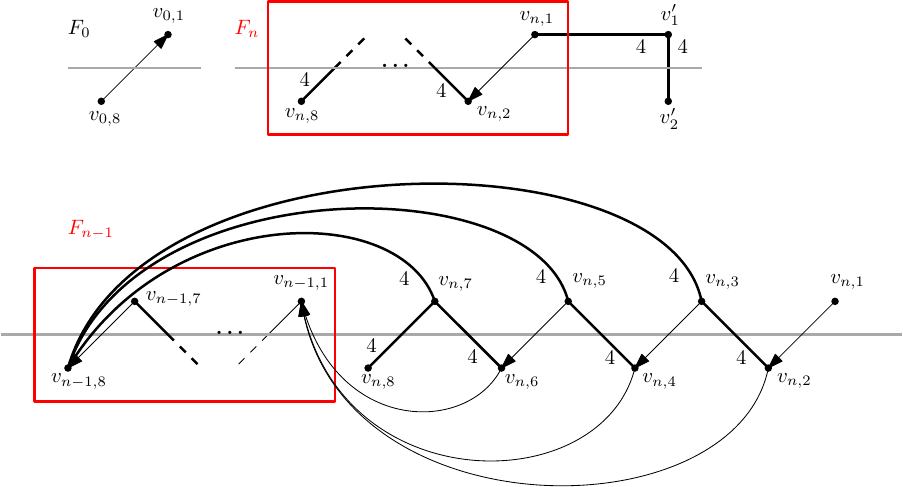}
    \caption{A valid $(k-1)$-labeling~$L$ and its gadget arrangement, for $k \geq 9$.
    Only labels~$L(\cdot,\cdot)$ with value other than two are specified.
    All labels to vertices have suitable values to ensure the label sum at each vertex is $k-1$. 
    Bold edges indicate simple gadgets, and the directed edges are flexible gadgets. The arrows indicate the direction of the forcing rule in \cref{lem:forcing_a} (i.e., if the gadget equipped to a directed edge is strict at the tail, then it is forced to be strict at the head).}
    \label{fig:michel_scott_2}
\end{figure}

We use the reduction from \cref{sec:reduction}, with some specialization.
Firstly, for the Max-Cut/Flip instance, we use the graph $H_n$ as described in Section~\ref{sec:michel_scott} as the graph $H$, for $n \geq 1$, and we use the corresponding weight and initial cut as described in the section.
Secondly, we use the following labeling~$L$:
(See \cref{fig:michel_scott_2} for a depiction of the labeling and gadget assignment.) 
\begin{itemize}
    \item For $i \in \{1, \dots, n\}$ and $q \in \{3, 5, 7\}$, we have $L(v_{i,q}, v_{i,q}v_{i-1,8}) = 4$;
    \item For $i \in \{1, \dots, n\}$ and $q \in \{2, 4, 6\}$, we have $L(v_{i,q}, v_{i,q}v_{i,q+1}) = 4$;
    \item $L(v_{n,8}, v_{n,8}v_{n,7}) = 4$;
    \item $L(v'_1, v_{n,1}v'_1) = L(v'_1, v'_{1}v'_2) = 4$;
    \item $L(x, z) = 2$, for all other pairs of an $H$-vertex $x$ and an $H$-edge $z$ not mentioned above and
    \item $L(v_{n,8}) = L(v_{n,1}) = k-5$; $L(v'_{2}) = k-3$;
    for other $H$-vertex~$x$, we have $L(x) = k-9$. (As $k \geq 9$, these labels are nonnegative.)
\end{itemize}
For every $H$-edge $xy$, if $L(x,xy) = L(y,xy) = 2$, we equip the flexible gadget to $xy$.
Otherwise, we use the $(4,2)$-simple gadget as described in \cref{lem:simple_gadgets} instead.

Observe that the labeling above is a valid $(k-1)$-labeling.
Further, consider the subgraph of $H$ containing all $H$-edges corresponding to the flexible gadgets.
This subgraph is a forest with the leaves: $v_{0,8}$, $v_{n,1}$, $v_{n,3}$, $v_{n,5}$, $v_{j,3}$, $v_{j,5}$, $v_{j,7}$, and $v_{j,8}$ for $j \in \{1, \dots, n-1\}$. 
By \cref{lem:strict_flexible_gadgets}, all parity gadgets are strict.

Then \cref{lem:all_exp_9} follows from \cref{thm:all_exp_maxcut} and \cref{lem:correspondence}.

\section{All-exp property for $k \geq 5$}
\label{sec:all-exp-5}
In this section, we prove \cref{thm:all_exp_5} that asserts the all-exp property of TSP/\kopt for $k \geq 5$.
This proof is similar to that of \cref{lem:all_exp_9} with a few changes.
Firstly, we modify the Michel-Scott construction for the Max-Cut/Flip instance.
In particular, we replace certain edges by paths of odd length.
Secondly, we introduce a new gadget, called the \emph{double gadget} that simulates two adjacent edges simultaneously.
Lastly, we do not insist that all gadgets are strict.
However, we argue that with our chosen initial tour, we cannot encounter a subtour other than subtours (1)-(4) in any parity gadget by a $k$-swap sequence.

\subsection{Modified Michel-Scott construction}
\label{sec:michel_scott_modified}
\begin{figure}
    \centering
    \includegraphics[page=3]{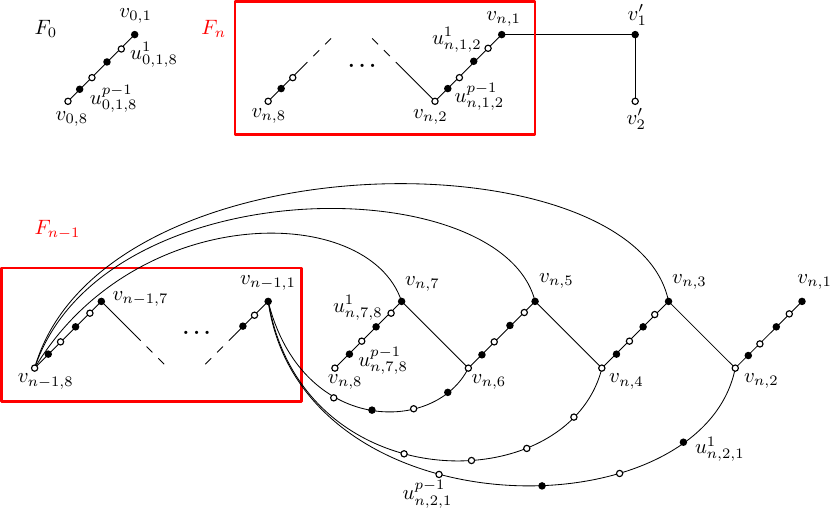}
    \caption{Modified Michel-Scott construction, where we replace certain edges in \cref{fig:michel_scott_1} with paths of length $p$. Here, we indicate the sets of the initial cut by the colors of the vertices.}
    \label{fig:michel_scott_3}
\end{figure}

See \cref{fig:michel_scott_3} for a depiction of the modification explained in this section.
Recall the construction by Michel and Scott~\cite{Michel_Max_Cut_4} and the unique maximal improving flip sequence~$L_n$ in Section~\ref{sec:michel_scott}.
Let $p$ be an odd number that is at least three.
We observe that for any consecutive pair $(v, v')$ in the sequence $L_n$, $vv'$ is an edge in $H_n$ and $(v', v)$ is not a contiguous subsequence of $L_n$.
In that case, for such consecutive pair $(v, v')$, we orient the edge $vv'$ in $H_n$ from $v$ to $v'$.
Then we obtain a partial orientation $\overrightarrow{H}_n$ of $H_n$.

For $i \in \{0, \dots, n\}$ and $q \in \{1, \dots, 7\}$, note that the vertex~$v_{i,q}$ only has one 
out-neighbor~$v_{i',q'}$ in $\overrightarrow{H}_n$, for some $i', q'$. 
We replace the edge $v_{i,q}v_{i',q'}$ by a path of length~$p$ $(v_{i,q}, u^{1}_{i,q,q'}, \dots, u^{p-1}_{i,q,q'}, \allowbreak v_{i',q'})$, with $p-1$ new vertices $u^{1}_{i,q,q'}, \dots, u^{p-1}_{i,q,q'}$.
The weights of the new edges and which set of the cut the new vertices belong to depend on the sign of the weight of the original edges.
In particular, let $\omega$ be the original weight of $v_{i,q}v_{i',q'}$ and $\varepsilon$ be a very small number (say, $2^{-n}$), and define $u^0_{i,q,q'} := v_{i,q}$ and $u^p_{i,q,q'} := v_{i',q'}$.
If $\omega > 0$, then we assign the weights for the edges along the path in decreasing order: for $j = 0, \dots, p-1$, the edge $u^{j}_{i,q,q'} u^{j+1}_{i,q,q'}$ has weight $\omega - j \varepsilon$.
Further, we assign $u^j_{i,q,q'}$ to the same set of the cut as $v_{i,q}$ for $j$ even, and to that as $v_{i',q'}$ for $j$ odd.
If $\omega < 0$, we assign the weights for the edges along the path in increasing order: for $j = 0, \dots, p-1$, the edge $u^{j}_{i,q,q'} u^{j+1}_{i,q,q'}$ has weight $\omega + j \varepsilon$.
Further, we assign $u^j_{i,q,q'}$ to the same set of the cut as $v_{i,q}$, for all $j \in [p-1]$.
The resulting (undirected) graph, weight, and cut after all replacements are the graph $H$, the weight $w$, and the initial cut~$\chi_0$ that we will use for the reduction.

We show that the uniqueness property carries over to the new instance.
\begin{lemma}
    There is a unique maximal improving flip sequence~$L'_n$ from the cut~$\chi_0$ of $H$.
\end{lemma}
\begin{proof}
    Let $L_n$ be the unique maximal improving flip sequence of the original instance $H_n$.
    For every consecutive pair $(v_{i,q}, v_{i',q'})$ of $L_n$ such that $i \in \{0, \dots, n\}$ and $q \in \{1, \dots, 7\}$, we replace the pair by the sequence $(v_{i,q}, u^{1}_{i,q,q'}, \dots, u^{p-1}_{i,q,q'}, v_{i',q'})$.
    Let $L'_n$ be the resulting sequence after all replacements.

    We say a vertex is \emph{flippable}, if a flip of that vertex is improving.
    A \emph{flip profile} of a vertex~$v$ is a set of its neighbors, such that if these neighbors are the only neighbors of $v$ that are in the same set as $v$ in the cut, then $v$ is flippable.
    
    Observe that along a replacement path $(v_{i,q}, u^{1}_{i,q,q'}, \dots, u^{p-1}_{i,q,q'}, v_{i',q'})$, if the weight $\omega$ of the original edge $v_{i,q}v_{i',q'}$ is positive, the weight assignment ensures that for $j = 1, \dots, p-1$, the vertex $u^j_{i,q,q'}$ is only flippable, when it is in the same set as $u^{j-1}_{i,q,q'}$ (i.e., it does not matter which set contains $u^{j+1}_{i,q,q'}$).
    On the other hand, if $\omega < 0$, the weight assignment ensures that for $j = 1, \dots, p-1$, the vertex $u^j_{i,q,q'}$ is only flippable, when it is in the \emph{different} set as $u^{j-1}_{i,q,q'}$ (again, it does not matter which set contains $u^{j+1}_{i,q,q'}$).
    
    Additionally, we can establish the following one-to-one correspondence between a flip profile of a vertex $v$ in $H_n$ and that of $v$ in $H$. 
    In one direction, for every vertex~$v'$ in the flip profile of $v$ in $H_n$, we replace $v'$ by the vertex closest to $v$ in the shortest path connecting $v$ and $v'$ in $H$.
    In the other direction, for a vertex $v''$ in the flip profile of $v$ in $H$, if $v''$ is not a vertex in $H_n$, then we follow the path starting from $v$ and contain the edge $vv''$ until we reach a vertex $v'$ that also appears in $H_n$. 
    We then replace $v''$ by $v'$ in the flip profile of $v$.
    Since we choose $\varepsilon$ to be very small, it is easy to check that the above constitutes a one-to-one correspondence.

    Combining the above observations with the fact that we only replace the outgoing edges of vertices with out-degree one in the partial orientation~$\overrightarrow{H}_n$, it is easy to verify from the uniqueness of $L_n$ that $L'_n$ is a unique maximal improving flip sequence of $H$ from~$\chi_0$.
\end{proof}

\subsection{Double gadget}
Let $xy$ and $xt$ be two $H$-edges.
Denote $z := xy$.
A \emph{double gadget} replaces a gateway $YY'$ of $y$, a gateway $TT'$ of $t$, the $z$-edge $ZZ'$, and a subpath $(X, X_1, X_2, X')$ of the second-set path of $x$, where $XX_1$ and $X_2X'$ are two gateways of $x$.
Note that the vertices $X_1$ and $X_2$ are removed from the graph $G$, when we equip the double gadget to the pair $xy$ and $xt$.
Further, the $xt$-edge is not replaced by any gadget. 
As the third edge in a path of length five in $G$, the $xt$-edge is then used in every tour of $G$.
We define the external and internal edges of the double gadget similar to those of parity gadgets.

\begin{figure}[t!]
    \centering
    \includegraphics[width=\textwidth]{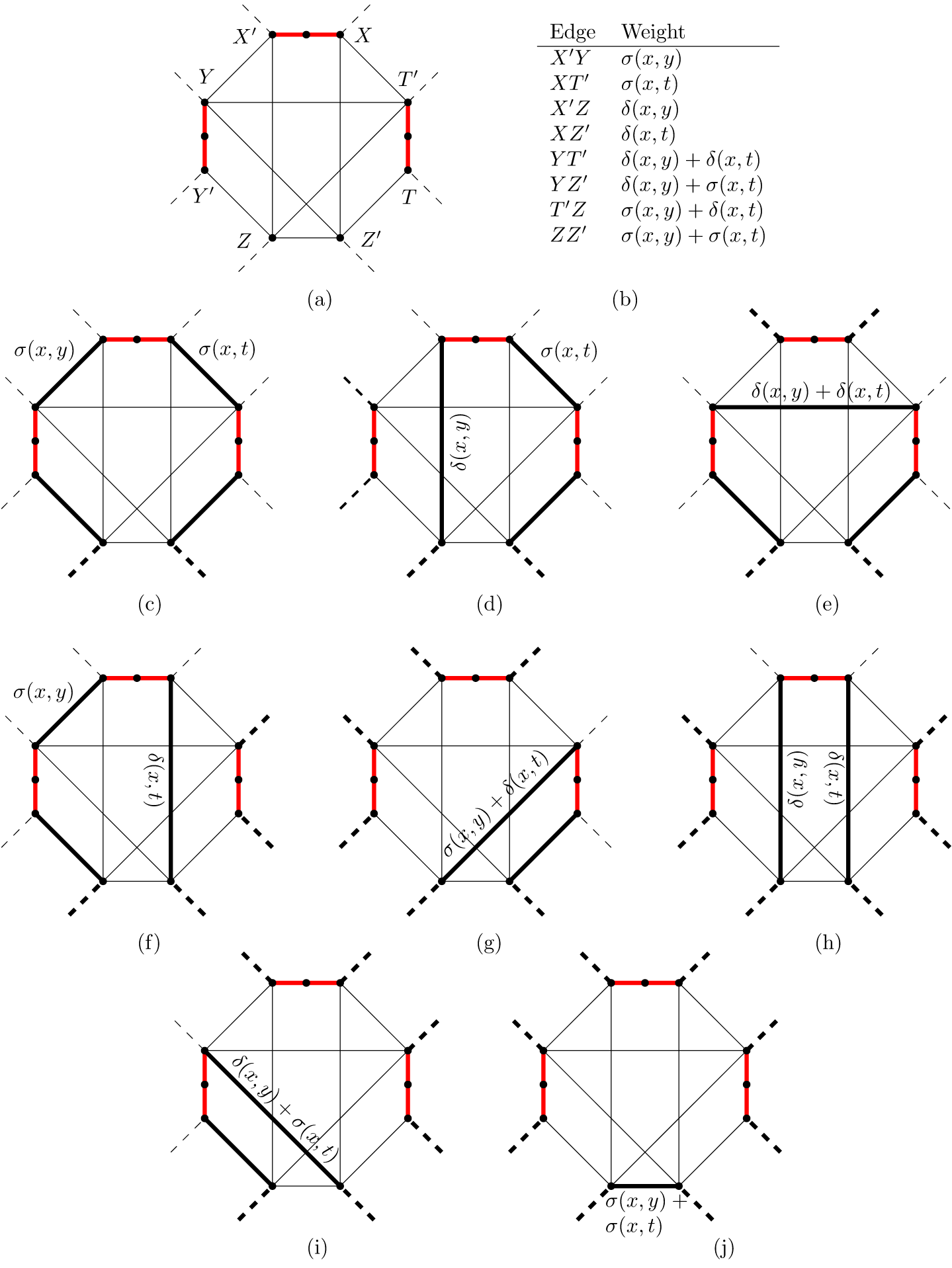}
    \caption{A (2,2,2)-double gadget (a) and the listing of edges with nonzero weights (b). The other panels show the eight subtours when the gadget is locally strict, including the nonzero weights of the edges in the subtours.
    }
    \label{fig:double_gadget}
\end{figure}

A double gadget has to guarantee at least eight possible subtours (with subtours defined analogously to subtours defined in \cref{def:parity-gadget}).
A $\{Z, Z'\}$-path is always present in these subtours.
The eight subtours corresponds to all possibilities of containing an $\{X, X'\}$-path, a $\{Y, Y'\}$-path, or a $\{T, T'\}$-path.

Let $\sigma(x,y)$ and $\delta(x,y)$ be the same-set and different-set weights for the edge $xy$, respectively (i.e., they correspond to the weights when $x$ and $y$ are in the same set and in different sets).
The numbers $\sigma(x,t)$ and $\delta(x,t)$ are defined analogously.
For each of the eight subtours, the total weight of the internal edges in the subtour is the sum of two numbers, $a_{xy}$ and $a_{xt}$.
$a_{xy}$ takes value $\sigma(x,y)$ if an $\{X, X'\}$-path and a $\{Y, Y'\}$-path are both present or both absent; and it takes value $\delta(x,y)$ if exactly one of these paths is present.
$a_{xt}$ is defined analogously.

We extend the definition of an $(r_x, r_y)$-parity gadget to an $(r_x, r_y, r_t)$-double gadget
in the obvious way, e.g., $r_x$ is the number of internal edges that have to be removed if in a subtour an $\{X,X'\}$-path is added. 

In the following proof, we use the (2,2,2)-double gadget as depicted in \cref{fig:double_gadget}.
One can easily verify that the graph is indeed a (2,2,2)-double gadget.
Note that the gadget also allows subtours other than those shown in the figure.
However, we will show later that these other subtours do not appear in the improving swap sequence of concern.

\subsection{Proof of \cref{thm:all_exp_5}}
We use the reduction described in \cref{sec:reduction}, from the Max-Cut instance $(H, w)$ and the initial cut $\chi_0$ indicated in \cref{sec:michel_scott_modified}.
In the construction of the Max-Cut instance, we use a constant $p$ which is odd and more than $2k$, where we recall that $p$ is the length of the paths that replace certain edges.

We redefine a labeling as follows: For each $H$-edge $xy$, either it is \emph{singly labeled} (i.e., we assign an integer label to the tuples $(x, xy)$ and $(y, xy)$) or it is \emph{jointly labeled} with another edge $xt$ (i.e., we assign an integer label to each of $(x, \{xy, xt\})$, $(y, \{xy, xt\})$, and $(t, \{xy, xt\})$).
We also assign a nonnegative integer label to each $H$-vertex individually.
A labeling~$L$ is then \emph{valid}, if there exists an $(L(x, xy), \allowbreak L(y, xy))$-parity gadget for each singly labeled $H$-edge~$xy$, and there exists an $(L(x,\{xy,xt\}), \allowbreak L(y, \{xy, xt\}), L(t, \{xy, xt\}))$-double gadget for each jointly labeled pair $xy$ and $xt$.

\begin{figure}[t!]
    \centering
    \includegraphics[width=\textwidth, page=4]{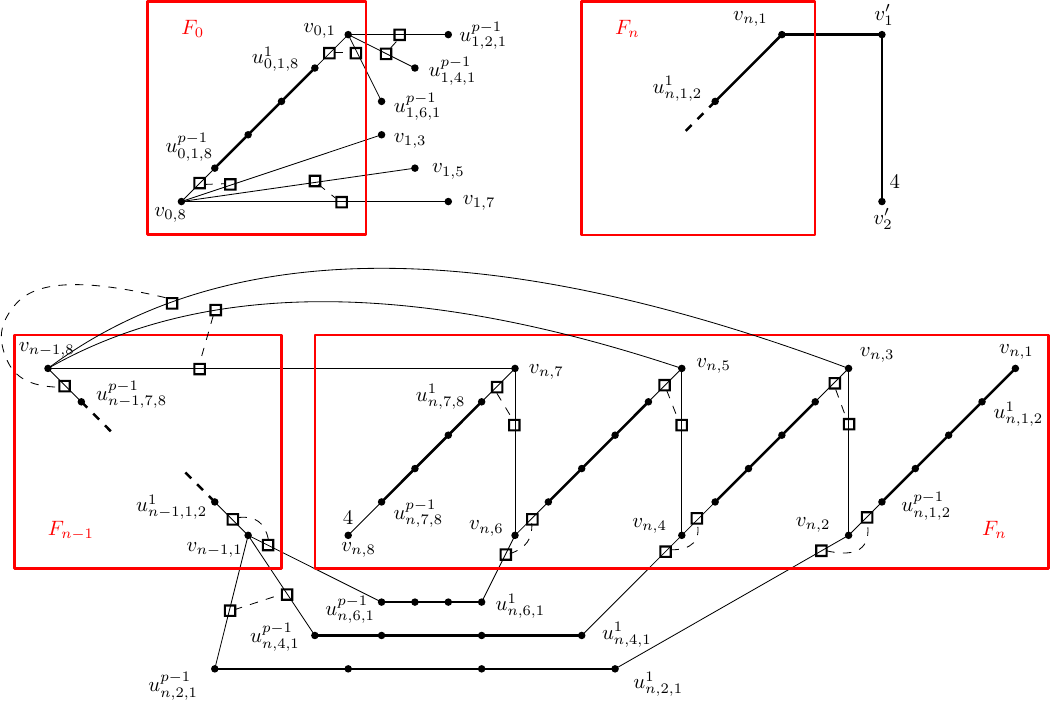}
    \caption{A valid 4-labeling. Each $H$-vertex has label zero. Other unspecified labels have value two. Bold edges are singly labeled, while pairs of edges indicated by connected pairs of boxes are jointly labeled.
    }
    \label{fig:michel_scott_4}
\end{figure}

Then we jointly label these pairs of edges:
\begin{itemize}
    \item $\{v_{0,8}v_{1,7}, \, v_{0,8}v_{1,5}\}$ and $\{v_{0,8}v_{1,3}, \, v_{0,8}u^{p-1}_{0,1,8}\}$;
    \item $\{v_{0,1}u^{p-1}_{1,2,1}, \, v_{0,1}u^{p-1}_{1,4,1}\}$ and $\{v_{0,1}u^{p-1}_{1,6,1}, \, v_{0,1}u^1_{0,1,8}\}$;
    \item $\{v_{i-1,8}v_{i,7}, \, v_{i-1,8}v_{i,5}\}$ and $\{v_{i-1,8}v_{i,3}, \, v_{i-1,8}u^{p-1}_{i-1,7,8}\}$, for $i \in \{2, \dots, n\}$;
    \item $\{v_{i-1,1}u^{p-1}_{i,2,1}, \, v_{i-1,1}u^{p-1}_{i,4,1}\}$ and $\{v_{i-1,1}u^{p-1}_{i,6,1}, \, v_{i-1,1}u^1_{i-1,1,2}\}$, for $i \in \{2, \dots, n\}$;
    \item $\{v_{i,q}u^{1}_{i,q,q+1}, \, v_{i,q}v_{i,q-1}\}$, for $i \in \{1, \dots, n\}$ and $q \in \{3, 5, 7\}$; and
    \item $\{v_{i,q}u^{p-1}_{i,q-1,q}, \, v_{i,q}u^1_{i,q,1}\}$, for $i \in \{1, \dots, n\}$ and $q \in \{2, 4, 6\}$.
\end{itemize}
The remaining edges are singly labeled.
We set $L(x) = k-5$ for each $H$-vertex~$x$.
Moreover, we define 
$L(v_{n,8}, \,v_{n,8}u^{p-1}_{n,7,8}) = L(v'_2, v'_1 v'_2) = 4$, and all other labels get value two.
See \cref{fig:michel_scott_4} for an illustration.

Next, we specify the corresponding gadget arrangement.
We assign the XOR gadget of order~$L(x)$ to each $H$-vertex~$x$.
Each jointly labeled pair of edges is equipped with the (2,2,2)-double gadget.
Each singly labeled edge with two incident labels of two is equipped with the flexible gadget.
Lastly, the remaining singly labeled edges (namely, $v_{n,8}u^{p-1}_{n,7,8}$ and $v'_1 v'_2$) are equipped with the 
$(4,2)$-simple gadget from \cref{lem:simple_gadgets}. 
We then obtain a valid $(k-1)$-labeling.

Next, we argue that although the flexible gadget and the (2,2,2)-double gadget allow many subtours, we only encounter subtours (1)-(4) for the flexible gadget and the subtours in \cref{fig:double_gadget} for the double gadget.
For this, we first define another concept related to strictness.
Note that strictness is defined to be a ``global" property; that is, it holds for all the possible tours in the graph~$G$.
Now we define local strictness that holds for a specific tour.
Particularly, for a parity or double gadget related to an $H$-vertex~$x$, it is \emph{locally strict at $x$} with respect to a tour, if its two external edges that are also doors of $x$ are both either in the tour or not in the tour.
If a gadget is locally strict at all its related vertices with respect to a tour, then we say the gadget is \emph{locally strict} with respect to that tour.
Note that the eight subtours shown in \cref{fig:double_gadget} are all possible subtours of the (2,2,2)-double gadget, if we require the double gadget to be locally strict. 
Then we have the following lemma.

\begin{lemma}
\label{lemma:k-swap-locally-strict}
    Suppose all gadgets are locally strict with respect to a tour~$\tau$. 
    Then with a $k$-swap from $\tau$, we cannot obtain a tour~$\tau'$, such that there exists a gadget that is not locally strict with respect to~$\tau'$.
\end{lemma}
\begin{proof}
    Suppose for the sake of contradiction, we obtain a tour~$\tau'$ by a $k$-swap from $\tau$, such that there exists a gadget that is not locally strict at a related $H$-vertex with respect to~$\tau'$.
    For the remainder of the proof, unless otherwise stated, all references of local strictness are with respect to the tour~$\tau'$.
    Further, unless specified otherwise, a gadget referred below is either a flexible gadget or a double gadget.

    The $H$-vertices $v_{n,8}$ and $v'_2$ are related to only simple gadgets, which are always (locally) strict.
    Each other $H$-vertex is related to exactly two gadgets.
    We have the following observations for these $H$-vertices:
    \begin{itemize}
        \item[(a)] For an $H$-vertex $x$ that is related to two gadgets, either both gadgets are locally strict at $x$ or both gadgets are not locally strict at $x$. 
        \item[(b)] Each gadget is either locally strict at all its related $H$-vertices or not locally strict at exactly two related $H$-vertices.
    \end{itemize}

    To see (a), note that the second-set path of $x$ has exactly three surviving doors (i.e., the doors that are not removed by any double gadget):
    $x_{\ell} x_1$, the closest door to $x_r$, and $x'_a x_b$ for some $a, b \in \{1, \dots, 4\}$.
    Since the first two doors are both either in the tour or not in the tour by \cref{lem:end_doors}, the two gadgets related to $x$ are either both locally strict or both not locally strict at $x$. Note that we can apply \cref{lem:end_doors} here as strictness is not used in the proof of \cref{lem:end_doors}.
    
    To see (b), note that for the flexible and (2,2,2)-double gadgets, the number of external edges on a tour has to be even.
    Hence, if a gadget is not locally strict at an $H$-vertex, then it has to be not locally strict at exactly one other related $H$-vertex.

    Using the two observations above, we obtain a sequence $S := (x^0, g^0, x^1, g^1, \dots)$, where for $i \geq 0$, $g^i$ is not locally strict at $x^i$ and $x^{i+1}$.
    Further, we insist that each element of the sequence only appears once.
    By the observations (a) and (b), the sequence~$S$ must end at a gadget $g^{j}$, where $g^j$ is not locally strict at $x^0$.

    Let $S'$ be the sequence obtained from $S$ by keeping only the $H$-vertices.
    In other words, $S' = (x^0, \dots, x^j)$.
    Observe that since all gadgets are locally strict with respect to the original tour $\tau$, for the two gadgets that are related to some $H$-vertex $x^i$, we need to add or remove at least one door in the second-set path of $x^i$.
    Moreover, this door is an external edge of the two gadgets related to~$x^i$.
    Therefore, in order to add or remove this door, we have to remove or add an adjacent internal edge in each of the related gadgets.
   
    This implies that the number of edges in the swap from $\tau$ to $\tau'$ is at least the length of $S'$.

    Next, we obtain another sequence $S''$ by adding certain $H$-vertices to $S'$.
    For $i \geq 1$, if $g^i$ is a double gadget and $x^{i}x^{i+1}$ is not an $H$-edge, then $g^i$ is related to another $H$-vertex $\tilde{x}^i$ that is adjacent to both $x^i$ and $x^{i+1}$ in $H$.
    Further, $g^i$ must be locally strict at $\tilde{x}^i$, since $g^i$ can only be not locally strict at at most two $H$-vertices.
    This implies that $\tilde{x}^i$ is not in the sequence $S$.
    In that case, we add $\tilde{x}^i$ in between $x^i$ and $x^{i+1}$ in the sequence $S'$.
    Observe that the final resulting sequence $S''$ has distinct vertices, and every consecutive pair in the cyclic order are adjacent in $H$.
    Therefore, $S''$ corresponds to a cycle in $H$.

    Finally, recall that in \cref{sec:michel_scott_modified}, we added paths of length $p$ into the graph $H_n$ in order to obtain $H$.
    If we remove these paths from $H$, then we obtain a forest.
    This means that each cycle in $H$ has to contain one of these paths of length $p$.
    Since $p > 2k$, we conclude that every cycle in $H$ has length more than $2k$.

    In summary, the number of edges in the swap from $\tau$ to $\tau'$ is at least $|S'|$, which is at least $|S''|/2$. 
    Further, $S''$ is a cycle in $H$, and all cycles in $H$ have length more than $2k$.
    It follows that we cannot obtain $\tau'$ from $\tau$ by a $k$-swap, a contradiction.
\end{proof}

The above lemma implies that since all gadgets are locally strict with respect to the original tour, this also applies to all tours in any improving $k$-swap sequence from this original tour.
With this observation, we can now follow the same arguments as in \cref{sec:reduction},
as long as we restrict them to tours in which all gadgets are locally strict (e.g. instead of ``any tour”
we have ``any tour in which all gadgets are locally strict” in \cref{lem:tour_edges}).
Note that the tours corresponding to cuts are those that are locally strict.
Hence, a correspondence between an improving flip sequence and an improving $k$-swap sequence as in \cref{lem:correspondence} still holds if we use the following slight variation that can be proved analogously:

\begin{lemma}
    Let $I$ be a Max-Cut/Flip instance and $\sigma$ be an initial cut for $I$.
    Suppose for some $s$, there is a valid $s$-labeling~$L$ for $I$ such that 
    \cref{lemma:k-swap-locally-strict} holds for $k = s + 1$. 
    Then we can reduce $I$ to a TSP/\kopt instance~$I'$, for $k = s+1$, and obtain an initial tour~$\tau$ 
    (in which all gadgets are locally strict)
    from~$\sigma$, such that there is a one-to-one correspondence between improving flip sequences for $I$ and $\sigma$ and improving $k$-swap sequences for $I'$ and $\tau$.
\end{lemma}

Since we have a valid $(k-1)$-labeling and because of \cref{lemma:k-swap-locally-strict} the theorem then follows.

\section{\pls-Completeness for $k \geq 17$}
\label{sec:PLS-complete-proof}
In this section, we prove the PLS-completeness of TSP/\kopt for $k \geq 17$. With this result we not only
improve the value $k \ge 14,208$ from Krentel~\cite{Kre1989,HH2023}. We also present the first rigorous PLS-completeness proof for TSP/\kopt as Krentel's proof has a substantial gap. He assumes without proof that 
no edges of infinite weight can appear 
in a local optimum. The definition of PLS-completeness~(see \cref{def:pls_complete}) requires that
the function $g$ maps local optima to local optima. Therefore, one either has to show that a local optimum cannot contain edges of infinite weight. 
Or one has to show how to extend the definition of the function $g$ to local optima that contain edges of infinite weight. Both are not done in the paper of Krentel~\cite{Kre1989} and there is no obvious way how to fill this gap. 

For our reduction we can prove in~\cref{lem:no_infty_local_optima} below that local optima cannot contain edges of infinite weight. There seems not to be a generic way to prove such a result for arbitrary
TSP instances as for example those constructed by Krentel~\cite{Kre1989}.
A result similar to~\cref{lem:no_infty_local_optima} was obtained by Papadimitriou~\cite{pap1992} 
for the Lin-Kernighan heuristic.

\plscomplete*

Our proof of \cref{thm:PLS_complete_17} follows closely the proof of \cref{lem:all_exp_13}.
However, there are three key differences.
Firstly, while the all-exp property is known to hold for Max-Cut instances with maximum degree four (\cref{thm:all_exp_maxcut}),
the PLS-completeness of Max-Cut/Flip is only known for maximum degree five: 

\begin{theorem}[\cite{Elsaesser_Max_Cut_5}]
\label{thm:PLS_complete_17_max_cut}
    Max-Cut/Flip is \pls-complete, even when restricted to graphs of maximum degree five.
\end{theorem}

Secondly, we impose certain structure on the graph $G$ in the reduction.
Particularly, we specify which gateways a parity gadget can replace.

Lastly, 
recall that the TSP instance requires a complete graph $G_{\infty}$, which we obtain from $G$ by adding the missing edges, which we also call the \emph{non-edges}.
By choosing suitable weights for the non-edges we will be able to prove in~\cref{lem:no_infty_local_optima} that no locally optimal tour of $G_{\infty}$ can contain a non-edge.

We first show the proof of \cref{thm:PLS_complete_17}.
The proof hinges on a technical lemma (\cref{lem:no_infty_local_optima}), whose proof we present separately in \cref{sec:proof_lem_no_infty}.
In turn, this lemma relies on a few structural observations of $G$ and the (4,2)-simple gadget, which we discuss in \cref{sec:prelim_lem_no_infty}.

\begin{proof}[Proof of \cref{thm:PLS_complete_17}]
    We use the reduction from Max-Cut to TSP as described in \cref{sec:reduction}.
    By \cref{thm:PLS_complete_17_max_cut}, we can assume in the Max-Cut instance $(H,w)$, that $H$ is a graph of maximum degree five.

    We assign an orientation on the $H$-edges, by repeating the following procedure: Until all $H$-edges have an orientation, we take a maximal (possibly closed) walk in the subgraph of unoriented $H$-edges, and we orient the edges along the walk.
    Observe that each degree-five vertex then has in-degree at most three.
    For every directed $H$-edge $z$ with head $x$ and tail $y$, we label $(x,z)$ and $(y,z)$ with four and two, respectively.
    Next, for each $H$-vertex~$x$, we assign an integer label to $x$, such that the label sum at~$x$ is $k-1$. (Note that this label is nonnegative, as $k \geq 17$.)
    Corresponding to these labels we use the $(4,2)$-simple gadgets from \cref{lem:simple_gadgets} and the XOR-gadgets.
    Hence, this is a valid $(k-1)$-labeling.
    We denote it by $L$.
    
    Next, we specify the gadget arrangement corresponding to $L$ as follows.
    Recall that $n$ is the number of $H$-vertices.
    Let $x^1, \dots, x^n$ be the $H$-vertices. 
    For every $H$-vertex~$x$, we assign the XOR gadget of order~$L(x)$ to $x$.
    Let $\psi$ be the increasing lexicographical order of the $H$-edges with respect to the $H$-vertex indices.
    That is, for $i < i'$ and $j < j'$, the $H$-edge $x^i x^{i'}$ precedes the $H$-edge $x^j x^{j'}$ in the order $\psi$, if either $i < j$ or $i = j$ and $i' < j'$.
    Then we equip the parity gadgets according to the labeling $L$, such that when we go along the second-set path for each $H$-vertex~$x$ from $x_{\ell}$ to $x_r$, the $H$-edges corresponding to the related gadgets appear in their order in $\psi$.
    We equip the gadgets such that for a (4,2)-simple gadget~$g$ related to an $H$-vertex~$t$, either the vertex~$X$ or~$Y$ in $g$ is adjacent to either $t_{\ell}$ or a vertex in a gadget that is also related to $t$ and precedes $g$ in $\psi$.
    With a slight abuse of notation, we also use $\psi$ to denote the order of the gadgets in the same order of the corresponding $H$-edges. We will rely on this ordering $\psi$ in \cref{sec:proof_lem_no_infty}.
    
    Let $G$ and $w$ be the resulting graph and edge weight function.
    We have the following lemma that we will prove in \cref{sec:proof_lem_no_infty}.
    \begin{lemma}
    \label{lem:no_infty_local_optima}
        For $k \ge 3$ there exists a complete graph $G_{\infty}$ and with a corresponding edge weight function $w'$ obtained from $G$ and $w$ by adding the missing edges with suitable weights, such that for the TSP/\kopt instance consisting of $(G_{\infty}, w')$, all locally optimal tours only contain edges of $G$.
    \end{lemma}
    
    For a given tour $\tau$ we can map it to a cut $\sigma$ as follows: For each $H$-vertex $x$ we
    put $x$ into the first set if $\tau$ uses the left first-set edge. 
    Otherwise we put $x$ into the second set. Assume we have a tour $\tau$ that is a local optimum but the 
    corresponding cut is not a local optimum. By \cref{lem:no_infty_local_optima} the tour $\tau$ contains
    only edges from $G$.  By \cref{lem:tour_edges} 
    every $H$-vertex in a tour is either in the first-set case or in the second-set case.
    Therefore, our mapping implies that we have property~(*). Now  \cref{lem:correspondence_13_1}
    implies that $\tau$ is not a local optimum, a contradiction.

    Then the theorem follows from \cref{thm:PLS_complete_17_max_cut}.
\end{proof}

\subsection{Structural properties of the gadgets and $G$}
\label{sec:prelim_lem_no_infty}
\begin{figure}
    \centering
    \includegraphics[page=2]{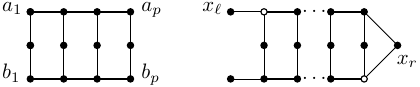}
    \caption{A standalone XOR gadget (left) and one assigned to an $H$-vertex~$x$ (right). Bold paths are rails. On the right, the top rail is the first rail, and the bottom the second rail. Their starts are colored white.}
    \label{fig:xor_gadget_2}
\end{figure}

We start by defining a few concepts related to the XOR gadget of order~$p$,
see \cref{fig:xor_gadget_2} for an illustration.
Recall from \cref{def:xor_gadget} that there are two paths of length~$p-1$: $(a_1, \dots, a_{p})$ and $(b_1, \dots, b_{p})$.
We call these paths the \emph{rails} of the XOR gadget.
For the XOR gadget that is assigned to an $H$-vertex~$x$, we define the \emph{first rail} of~$x$ to be the rail whose one endpoint is incident to the left first-set edge of $x$.
Its \emph{start} is this endpoint.
We call the other rail the \emph{second rail} of~$x$, and its \emph{start} is the endpoint that is adjacent to~$x_r$.

Next, in the $(4,2)$-simple gadget from \cref{lem:simple_gadgets}, there is an XOR gadget of order three with $\{X,X'\}$-path and $\{a,a'\}$-path as subtours.
We label the vertices along the two rails of the XOR gadget as $(X, b, a)$ and $(X', b', a')$.
(See \cref{fig:strict_gadget_1}.)

Then based on the construction of $G$ and the gadgets, we have the following observation on the degrees in $G$.

\begin{observation}
\label{lem:degrees}
    The following holds for the graph $G$:
    \begin{enumerate}[(a)]
        \item \label{item:all_degrees} A vertex of $G$ has degree two, three, or four.
        \item \label{item:adjacent_degree_two} Every vertex in $G$ is either a degree-two vertex or adjacent to one.
        \item \label{item:degree_four} Every degree-four vertex in $G$ is in a simple gadget.
        Every simple gadget in $G$ has four degree-four vertices (namely, $Z, Z', a$, and $Y'$). 
        \end{enumerate}
\end{observation}

\subsection{Proof of \cref{lem:no_infty_local_optima}}
\label{sec:proof_lem_no_infty}

Let $V_2$, $V_3$, and $V_4$ be the vertices in $G$ that have degree two, three, and four, respectively.
Further, denote $N_2 := |V_2|$, $N_3 := |V_3|$, $N_4 := |V_4|$, and $N = N_2 + N_3 + N_4$.
By \cref{lem:degrees}(\ref{item:all_degrees}), $(V_2, V_3, V_4)$ is a partition of the vertex set of $G$, and $N$ is the total number of vertices in $G$.
In this section, when we say degree-two, -three, -four vertices, the degree we refer to is the degree in $G$.

We now assign a priority $\phi(v)$ to every vertex~$v$ of $G$ in the decreasing order from $N$ to $1$, as follows.
Firstly, we assign the values in $\{N, \dots, N_3 + N_4 + 1\}$ to the vertices in $V_2$ in an arbitrary order.
Secondly, for $i = 1, \dots, n$, we assign the value of $N_3 + N_4 + 1 - i$ to $x^i_{\ell}$.
(Recall that $x^i_{\ell}$ and $x^i_r$ are the endpoints of the first-set edge of the $H$-vertex~$x^i$.)
Thirdly, for $i=1, \dots, n$, we first assign the next available priority values to the vertices along the first rail of $x^i$ from its start.
Then we assign the next value to $x^i_r$.
After that, we assign the next available priority values to the vertices along the second rail of $x^i$ from its start.
Lastly, for each simple gadget~$g$ in the order $\psi$, we assign the next available values of priority to all vertices in~$g$ in the following order $(Y, X, b, X', b', a', Z, Z', a, Y')$.
Note that we assign priorities to the degree-three vertices before assigning to the degree-four vertices in $g$.

Let $G_{\infty}$ be the complete graph obtained from $G$ by adding the \infedge{}s.
We call an edge of $G$ a \emph{\fedge}.
For convenience, we still use the word ``edge" to refer generally to either a \infedge or a \fedge.
In order to obtain the edge weight function $w'$ for $G_{\infty}$ from $w$, we prescribe the weights of non-edges, as follows. Let $M$ be the maximum edge weight appearing in $G$.
For a non-edge $vv'$, we assign $M \cdot 3^{\max\{\phi(v), \phi(v')\}}$ to be the weight of $vv'$.

Then it is easy to see that the following lemma immediately implies \cref{lem:no_infty_local_optima}.

\begin{lemma}
    \label{lem:swap_from_infty}
    Let $\tau$ be a tour of $(G_{\infty}, w')$ with a \infedge.
    Then there exists an improving $3$-swap for $\tau$.
\end{lemma}
\begin{proof}
    We define a tour $\tau^*$ containing only \fedges, as follows.
    The tour $\tau^*$ contains all incident edges of degree-two vertices.
    For an $H$-vertex $x$, if the left first-set edge of $x$ is in $\tau$, then $\tau^*$ contains the left and right first-set edge of $x$ and the subtour that is 
    incident to the left and right first-set edge
    of the XOR gadget assigned to $x$.
    Otherwise, $\tau^*$ contains all the doors of $x$ and the subtour that is 
    incident to the right second-set edge
    of the XOR gadget assigned to $x$.
    For an $H$-edge $xy$, in the corresponding gadget, we use the subtour (1) if the left first-set edges of $x$ and $y$ are in $\tau^*$, subtour (2) if the left first-set edge of $y$ is in $\tau^*$ but that of $x$ is not, subtour (3) if the left first-set edge of $x$ is in $\tau^*$ but that of $y$ is not, and subtour (4) if the left first-set edges of $x$ and $y$ are not in $\tau^*$.
    
    We say that a vertex of $G_{\infty}$ is \emph{$\tau^*$-consistent}, if the incident edges of that vertex in~$\tau$ are the same as those in $\tau^*$.
    Let $i$ be the smallest integer such that for any vertex $v$ of $G_{\infty}$ with $\phi(v) > i$, $v$ is $\tau^*$-consistent.
    Since $\tau$ has a \infedge, $i \geq 1$.
    Let $v$ be the vertex with $\phi(v) = i$.

    We start with the following claim.

    \begin{claim}
    \label{claim:3_swap_from_infty}
        Suppose there exist a \infedge $\tilde{v}v'$ in~$\tau$ and an edge $u\tilde{v}$ in~$\tau^*$ but not in~$\tau$.
        Further suppose that for any vertex~$v'' \neq u,\tilde{v}$, if $v''$ is not $\tau^*$-consistent, then $\phi(v'') < \phi(\tilde{v})$.
        Then there exists an improving 3-swap that removes $\tilde{v}v'$.
    \end{claim}
    \begin{claimproof}[Claim's proof]
        Fix an orientation of $\tau$.
        Let $u'$ be a vertex such that $uu'$ is in $\tau$ but not in $\tau^*$.
        This implies that $u'$ is not $\tau^*$-consistent, and hence, $\phi(u') < \phi(\tilde{v})$.
        Similarly, $\phi(v') < \phi(\tilde{v})$.
        If $(\tilde{v},v')$ and $(u,u')$ have the same orientation in $\tau$, then since $u\tilde{v}$ is a \fedge, and since $\phi(u')$ and $\phi(v')$ are smaller than $\phi(\tilde{v})$, replacing $\tilde{v}v'$ and $uu'$ by $u\tilde{v}$ and $u'v'$ yields a better tour.
        Otherwise, consider the path in $\tau$ between $u$ and $\tilde{v}$ without containing $u'$ and $v'$.
        If this path has only edges in $\tau^*$, then the path together with $u\tilde{v}$ forms a cycle in $\tau^*$ that is not $\tau^*$ itself (since $u'$ and $v'$ are not on the path), a contradiction to $\tau^*$ being a tour.
        Therefore, the path contains an edge $tt'$ that is not in $\tau^*$.
        This implies that $t$ and $t'$ are not $\tau^*$-consistent, and hence $\phi(t)$ and $\phi(t')$ are smaller than $\phi(\tilde{v})$. 
        By definition of $w'$ and the fact that $u\tilde{v}$ is an edge in $G$ we have that the weight of $\tilde{v}v'$ is by a factor~3 larger than the weights of each of $u\tilde{v}$, $t'u'$, and $tv'$ and therefore it is also larger than the sum of these three edge weights.
        Therefore, replacing $uu'$, $\tilde{v}v'$, and $tt'$ with $u\tilde{v}$, $t'u'$, and $tv'$ yields a better tour.
        Note that this argument (that the edge  $\tilde{v}v'$ has larger weight than the sum of the 
        weights of the edges $u\tilde{v}$, $t'u'$, and $tv'$)  also holds in case that $t$ or $t'$ coincide with $u$ or $\tilde v$.
    \end{claimproof}
    
    By \cref{claim:3_swap_from_infty}, if $v$ is incident to a \infedge, we have an improving 3-swap for $\tau$.
    Therefore, for the remainder of the proof, we assume that $v$ is incident to only \fedges.
    We consider the following five cases.
    
    \textbf{Case 1: $v \in V_{2}$.} 
    In this case, $v$ only has two incident \fedges, and they have to be both in $\tau^*$ and in $\tau$.
    Therefore, $v$ is $\tau^*$-consistent, a contradiction to the choice of $i$.

    \textbf{Case 2: $v$ has degree~$d\in\{3,4\}$ in $G$ and has at least $d-1$ neighbors in $G$ with priority higher than $i$.}
    By the choice of $i$, the neighbors with priority higher than $i$ are $\tau^*$-consistent.
    Hence, the edges that are incident to $v$ and one of these neighbors are either both in $\tau$ and $\tau^*$ or both not in $\tau$ and $\tau^*$.
    Hence, if the two incident edges of $v$ in $\tau^*$ are incident to two of these neighbors, then $v$ is $\tau^*$-consistent, a contradiction.
    Otherwise, $v$ is adjacent to exactly one of these neighbors in both $\tau$ and $\tau^*$.
    This implies the edge $vv'$ is in both $\tau$ and $\tau^*$, where $v'$ is the remaining neighbor of $v$ in $G$.
    However, this also means that $v$ is $\tau^*$-consistent, a contradiction.

    \textbf{Case 3: $v = x^j_{\ell}$ for some $H$-vertex~$x$ and $j \in \{1, \dots, n\}$.}
    Observe that $v$ is incident to the left first-set edge of~$x$, a door $vv'$ of $x$, and an edge $uv$, where $u \in V_{2}$.
    By the choice of $i$, $u$ is $\tau^*$-consistent, and hence, $uv$ is both in $\tau$ and $\tau^*$.
    By construction of $\tau^*$, if the left first-set edge of $x$ is in $\tau$, then it is also in $\tau^*$.
    Otherwise, $vv'$ is in $\tau^*$, and by the assumption that $v$ is incident to only \fedges, $vv'$ is also in $\tau$.
    Hence, $v$ is $\tau^*$-consistent, a contradiction.

    \textbf{Case 4: $v$ is in $V_3$ and is not covered by Cases 2 and 3.}
    By the construction of the gadget arrangement and the priority assignment using the order $\psi$, $v$ can only be the vertex $X'$ of some simple gadget~$g$.
    Then the two neighbors of $v$ that have degree more than two in $G$ are the vertex $b'$ of $g$ and a vertex $v'$ outside of the gadget~$g$.
    We consider two subcases.
    (See \cref{fig:case_3} for an illustration.)

    \begin{figure}[t!]
        \centering
        \includegraphics[scale=0.8]{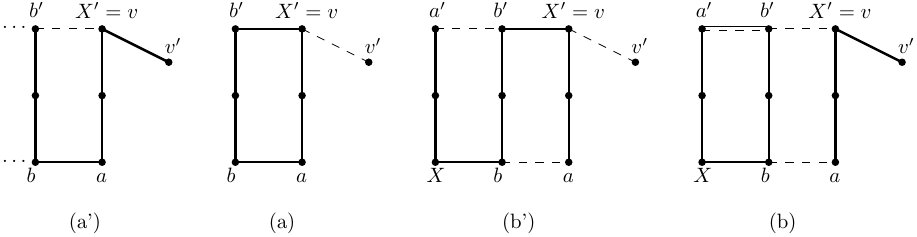}
        \caption{Illustration of Subcases 4a and 4b in the proof of \cref{lem:swap_from_infty}. 
        Bold edges are edges of $G$ present in a specific tour, while dashed edges are not present.
        Panels (a') and (b') corresponds to the tour $\tau^*$ in Subcases 4a and 4b, respectively.
        Since $X, b$, and $a$ are $\tau^*$-consistent, these panels imply the situations in the tour $\tau$ in panels (a) and (b).
        The edge $a'b'$ in panel (b) indicates that this edge may be in $\tau$ (which creates a 6-cycle) or not in $\tau$ (which means $b'$ is incident to a \infedge).}
        \label{fig:case_3}
    \end{figure}

    \textit{Subcase 4a: $vv'$ is in $\tau^*$ and $vb'$ is in $\tau$.}
    In this case, $ab$ has to be in $\tau^*$, by the design of the XOR gadget.
    By the priority assignment, $b$ has higher priority than $v$, and hence, $b$ is $\tau^*$-consistent.
    This implies that $ab$ is also in $\tau$.
    However, we then have a subgraph of $\tau$ that is the 6-cycle containing $a, b, b'$, and $v$, a contradiction to $\tau$ being a tour.
    
    \textit{Subcase 4b: $vv'$ is in $\tau$ and $vb'$ is in $\tau^*$.}
    In this case, $Xb$ has to be in $\tau^*$, by the design of the XOR gadget.
    With analogous argument as before, this edge is also in $\tau$, and hence $a'b'$ must not be in $\tau$, because otherwise, we have a 6-cycle.
    Since $b'$ is not adjacent to $v$ or $a'$ in $\tau$, it must be incident to a \infedge $b'u'$.
    Note that $\phi(b') = \phi(v) -1 = i-1$.
    Then applying \cref{claim:3_swap_from_infty} to the edges $b'u'$ and $b'v$, we obtain an improving 3-swap that removes $b'u'$.
    
    \textbf{Case 5: $v$ is in $V_4$ and is not covered by Case 2.}
    By \cref{lem:degrees}(\ref{item:degree_four}), $v$ is in some simple gadget $g$.
    By the choice of $i$ and by the priority assignment, if $v$ is not covered by Case 2, $v$ can only be the vertex $Z$ of $g$.    
    By \cref{lem:degrees}(\ref{item:adjacent_degree_two}), in $\tau^*$, $v$ is adjacent to a vertex $u \in V_2$.
    Using a similar argument as in Case~4, we deduce that $uv$ is also in $\tau$, and $v$ has a neighbor~$v'$ in $\tau^*$, such that $vv'$ is not in $\tau$.
    Let $t$ be the vertex such that $vt$ is the edge in $\tau$ but not in $\tau^*$.
    Then, as vertex $a'$ in $g$ has higher priority than $Z$, we only have to consider two subcases.

    \textit{Subcase 5a: $v'$ is $Y'$, and $t$ is $Z'$.}
    In this case, we have subtour~(2) in $\tau^*$. (See \cref{fig:strict_gadget_1}(c).)
    However, in $\tau^*$, $Z'$ should then be adjacent to $Y$ and a degree-two vertex, both of which have priority higher than $v$.
    Hence, $Z'$ is also adjacent to these two vertices in $\tau$.
    Then $Z'$ is adjacent to three vertices in $\tau$: $v$, $Y$, and the degree-two vertex, a contradiction to $\tau$ being a tour.

    \textit{Subcase 5b: $v'$ is $Z'$, and $t$ is $Y'$.}
    In this case, we have subtour~(4) in $\tau^*$.
    (See \cref{fig:strict_gadget_1}(e).)
    Observe that in this subtour, the neighbors of $a$ have priority higher than $v$.
    Hence, the incident edges of $a$ in $\tau$ and $\tau^*$ are the same.
    Accordingly, the edge $Z'a$ cannot be in $\tau$.
    Since the edge $ZZ'$ is not in $\tau$ either, and since the other neighbors of $Z'$ in $G$ other than $Z$ and $a$ have higher priority than $i$, we conclude that only one \fedge is incident to $Z'$ in $\tau$.
    This means $Z'$ is incident to a \infedge~$Z't'$ in $\tau$.
    Since $\phi(Z') = \phi(Z) - 1 = i-1$, and since $t'$ must not be $\tau^*$-consistent, the edges $Z't'$ and $Z'Z$ satisfy the condition of \cref{claim:3_swap_from_infty}.
    Hence, there is a 3-swap that removes $Z't'$.
    
    This completes the proof of the lemma.
\end{proof}

\section{Conclusion}

We have shown that for $k\ge 5$ the \kopt algorithm for TSP has the all-exp property, i.e.\ it has exponential running time 
for all possible pivot rules~(\cref{thm:all_exp_5}). Moreover, we proved that TSP/\kopt is 
PLS-complete for $k\ge 17$~(\cref{thm:PLS_complete_17}). In both cases we drastically lowered 
the so far best known value for $k$ which was $\gg 1000$. 
It was mentioned (without explaining the details) in~\cite{Kre1989} that there is a connection 
between the PLS-completeness of a problem and the all-exp property. This connection was made precise by Schäffer and Yannakakis~\cite{schaffer1991} who introduced the notion of \emph{tight} PLS-completeness. They proved that the
tight PLS-completeness of a problem implies the all-exp property. 
Our PLS-completeness result for TSP/\kopt relies on the PLS-completeness of Max-Cut/Flip for graphs
with maximum degree five~\cite{Elsaesser_Max_Cut_5}. 
As for the latter the tight PLS-completeness is not known we do not get 
tight PLS-completeness for TSP/\kopt. 

We put some effort into getting the constant in \cref{thm:all_exp_5} as small as possible.
In contrast, the constant in \cref{thm:PLS_complete_17} very likely can be lowered to $15$ by using our techniques from \cref{sec:all-exp-5}. However, this would require substantially more involved proofs. Finally we would like to
mention (as already observed by Krentel~\cite{Kre1989}) that our results also hold for \emph{metric} TSP as one can add a sufficiently large constant to all edge weights.

\paragraph{Acknowledgement.} 
This work was initiated at the ``Discrete Optimization'' trimester program of the Hausdorff Institute of Mathematics, University of Bonn, Germany in 2021.
We would like to thank the organizers for providing excellent working conditions and inspiring atmosphere. 

\bibliographystyle{alpha}
\bibliography{refs}

\end{document}